\documentclass[journal,10pt]{IEEEtran}

\usepackage{amssymb, amsmath, amsthm, subfigure}
\usepackage{amsfonts, bbm}
\usepackage{graphicx}
\usepackage{epstopdf}
\usepackage{times}
\usepackage{float}
\usepackage{array}
\usepackage{arydshln}
\usepackage{bibentry}
\usepackage{booktabs}
\usepackage{cite}
\usepackage{url}
\usepackage[caption=false]{subfig}
\usepackage{color}
\usepackage[table]{xcolor}
\usepackage[normalem]{ulem}
\usepackage{textcomp}

\newtheorem{thm}{\itshape \indent Theorem}
\newtheorem{lem}{\itshape \indent Lemma}
\newtheorem{col}{\itshape \indent Corollary}
\theoremstyle{remark}

\DeclareMathOperator*{\Exp}{Exp}
\newcommand*{\Scale}[2][4]{\scalebox{#1}{$#2$}}%
\DeclareMathOperator\erf{erf}
\DeclareMathOperator\erfc{erfc}
\begin{document}

\title{Cognitive and Energy Harvesting-Based D2D Communication in Cellular Networks: Stochastic Geometry Modeling and Analysis}

\author{Ahmed Hamdi Sakr and Ekram Hossain
\thanks{A. H. Sakr and E. Hossain are with the Department of Electrical and Computer Engineering, University of Manitoba, Winnipeg, Canada (emails:  Ahmed.Sakr@umanitoba.ca, Ekram.Hossain@umanitoba.ca). 
}
}

\maketitle

\begin{abstract}
While cognitive radio enables spectrum-efficient wireless communication, radio frequency (RF) energy harvesting from ambient interference is an enabler for energy-efficient wireless communication. In this paper, we model and analyze cognitive and energy harvesting-based device-to-device (D2D) communication in cellular networks. The cognitive D2D transmitters  harvest energy from ambient interference and use one of the channels allocated to cellular users (in uplink or downlink), which is referred to as the D2D channel, to communicate with the corresponding receivers. We investigate two spectrum access  policies for  cellular communication in the uplink or downlink, namely, random spectrum access (RSA) policy and prioritized spectrum access (PSA) policy. In RSA,  any of the available channels  including the channel used by the D2D transmitters can be selected randomly for cellular communication, while in PSA  the D2D channel is used only when all of the other channels are occupied. A D2D transmitter can communicate successfully with its receiver only when it harvests enough energy to perform channel inversion toward the receiver, the D2D channel is free, and the signal-to-interference-plus-noise ratio ($\mathsf{SINR}$) at the receiver is above the required threshold; otherwise, an outage occurs for the D2D communication.  We use tools from stochastic geometry to evaluate the performance of the proposed communication system model with general path-loss exponent in terms of outage probability for D2D and cellular users. We show that energy harvesting can be a reliable alternative to power cognitive D2D transmitters while achieving acceptable performance. Under the same $\mathsf{SINR}$ outage requirements as for the non-cognitive case, cognitive channel access improves the outage probability for D2D users for both the spectrum access policies. When compared with the RSA policy,  the PSA policy provides a better performance to the D2D users. Also, using an uplink channel provides improved performance to the D2D users in dense networks when compared to a downlink channel. For cellular users, the PSA policy provides almost the same outage performance as the RSA policy.

\end{abstract}


\section{Introduction}
Harvesting energy from non-traditional sources such as ambient interference is emerging as an attractive solution to power low-energy wireless communication devices \cite{adnan, paradiso,bouchouicha}. On the other hand, in order to improve spectrum utilization and mitigate the scarcity of spectrum, innovative solutions such as cognitive radio~\cite{sawy2013mc} and device-to-device (D2D) technology have been recently proposed to underlay current cellular networks \cite{doppler}.  

We consider cognitive D2D communication underlaying a multi-channel cellular network where the D2D transmitters are able to use only the harvested RF energy from the ambient interference that results from the concurrent downlink and uplink transmissions by both the macro base stations (BSs) and cellular users. After harvesting sufficient energy, each D2D transmitter performs spectrum sensing to opportunistically access a predefined nonexclusive D2D channel.  In a multi-channel environment, we consider two different spectrum access policies for cellular communication (in the uplink/downlink to/from BS) to enable the coexistence of the cellular and D2D users. We use a statistical approach based on stochastic geometry~\cite{baccelli2010stochastic,sawysurvey} to model and evaluate the performance of the proposed system in terms of outage probabilities of D2D users as well as cellular users. Note that  the outage for D2D users may occur due to either  insufficient amount of harvested energy, unavailability of the channel, or signal-to-interference-plus-noise ratio ($\mathsf{SINR}$) at the receiver falling below the required threshold. For cellular users, the outage occurs due to either unavailability of channels or insufficient $\mathsf{SINR}$. Due to their analytical tractability, we use independent Poisson Point Processes (PPPs) to model the locations of BSs, cellular users, and D2D users. The results from the analysis enable us to understand the impact of network parameters (such as base station (BS) and user density, number of channels, carrier/spectrum sensing threshold, and receiver sensitivity) on the performance measures and provide insightful guidelines for system design. 

\subsection{Related Work and Motivations}
In the context of energy harvesting in wireless networks, one way to evaluate the performance of the system under investigation is to use statistical modeling \cite{huang1, huang2, dusit, krik1}. Although statistical modeling gives insights into the long-term performance behavior and helps to select the statistically optimal network parameters, these parameters are not necessarily optimal on a short time-scale. On the other hand, tools from optimization theory can be used to model the network and evaluate the short-term performance to find optimal parameters that maximize certain objective functions \cite{rui1, rui2, derrick}. However, obtaining the optimal parameters on a short time basis generally increases the computational complexity and puts a burden on the system due to more frequent exchange of information. 

In \cite{huang1}, the authors deploy dedicated stations called ``power beacons'' that radiate out-of-band microwave signals to power all mobile devices. Under an outage constraint, the uplink cellular network performance is evaluated using a statistical model and the region of feasible operation is defined for different setups. In \cite{huang2}, the authors propose a cognitive radio model in which a low-energy secondary transmitter harvests RF energy from transmissions by primary users in its vicinity.  Statistical analysis is used to optimally choose network parameters such as power and density of secondary transmitters in order to maximize the spatial throughput while satisfying some outage constraints.  In \cite{dusit}, the authors use Ginibre determinantal point process to obtain bounds on the performance of a wireless sensor network with RF energy harvesting. The author in \cite{krik1} derives the outage probability and the average harvested energy of the simultaneous information and power transfer. The author considers a large-scale network with large number of randomly located transmitter-receiver pairs with and without relaying. 

In \cite{rui1}, the authors use dynamic programming to derive the optimal power control policy that minimizes the outage probability. The optimization problem is formulated and solved for block fading channels under energy harvesting constraints such that the transmit power is upper bounded by the amount of energy harvested. In \cite{rui2}, for a point-to-point wireless link, the authors assume that the receiver can either harvest energy from ambient RF signals or decode information at any point of time. For such a scenario, the authors use  optimization tools to obtain the optimal mode switching point that balances the tradeoff between the amount of harvested energy, data rate, and outage probability. In \cite{derrick}, the authors use fractional programming and dual decomposition to propose a resource allocation algorithm that maximizes the energy efficiency of a downlink single-cell orthogonal frequency division multiple access (OFDMA) network.

In the context of multi-channel cognitive cellular networks, the authors in \cite{sawy2013mc} provide a framework to model such a network in which macro BSs are underlaid with cognitive femtocell BSs. For the network under consideration, statistical analysis is used to obtain a long-term optimal spectrum sensing threshold that minimizes the outage probability of the cognitive femtocell BSs. On the other hand, in the context of D2D communication, the authors in \cite{andrewsD2D} use statistical analysis to investigate the effect of distance-based mode selection and power control on the  outage performance in the uplink. Another statistical framework is presented in \cite{sawyD2D} that takes the quality of the links between D2D users and BSs into consideration in the mode selection phase, furthermore, it accounts for the maximum transmit power of users. For  performance evaluation of D2D transmissions underlaying a cellular network, the authors in \cite{154165, 6516565, 51651} consider different scenarios and optimization problems. In \cite{154165}, the authors propose a greedy algorithm to solve the resource allocation problem where the optimization problem is formulated as a mixed-integer non-linear program to maximize the sum-rate of both cellular and D2D under $\mathsf{SINR}$ constraints. The authors in \cite{6516565} consider a network scenario in which cellular and D2D users share the same resources. The system aims to maximize the network throughput via mode selection and power control while satisfying spectral efficiency and power constraints. For some special cases, the optimal solution is obtained either  in a closed-form or by searching a finite set. In \cite{51651}, the authors propose a joint resource block allocation and power control scheme to maximize the spectrum utilization while fulfilling some interference constraints and traffic demands of cellular and D2D users, respectively. 

\subsection{Contributions and Organization}

The contributions of the paper can be summarized as follows:

\begin{itemize}

\item Using tools from stochastic geometry, we provide a tractable analytical framework for statistical analysis of cognitive D2D communication\footnote{We use the term ``cognitive D2D communication" in the sense that spectrum sensing is performed at each D2D transmitter before transmission to make sure that the channel designated for D2D transmission is not being used for cellular communication in the uplink or downlink. Here cognition is with respect to the cellular BSs and cellular users only, which is similar to the concept of ``semi-cognitive'' spectrum access in~\cite{hesham-spaswin13}.} using energy  harvested from the ambient interference. For a general path-loss exponent, we derive simple and closed-form expressions for the probability of harvesting sufficient energy, the probability that the channel to be used by D2D users is free, the $\mathsf{SINR}$ outage probability for both D2D and cellular users, and the overall outage probability for D2D users. We discuss the different trade-offs in the system and show the effect of varying network parameters such as spectrum sensing threshold, densities of BSs and cellular users, number of available channels, and sensitivity of the receivers on the system performance.

\item While D2D users perform spectrum sensing-based transmission and a channel inversion power control, two different spectrum access policies are used for cellular communication, namely, random spectrum access (RSA) and prioritized spectrum access (PSA) policies. For each spectrum access policy, we analyze the performance of energy-harvesting D2D communication. We also show how cellular users are affected by the adopted spectrum access policy.

\item We consider both the cases when D2D transmissions take place in a channel assigned for downlink cellular transmissions or uplink cellular transmissions. We investigate the different scenarios to show when uplink channels are preferable to downlink channels and vice versa. More specifically, we obtain a closed-form expression for the  value of the BS density after which uplink channels should be used to achieve a better performance for D2D communication  when compared to using downlink channels. 

\item We show that provisioning of multiple channels can be used along with cognition by D2D users to protect the cellular transmissions. For the same network parameters and $\mathsf{SINR}$ outage requirements, we also show that the overall outage performance of D2D users is always superior with the PSA policy compared to the RSA policy while the performance of the cellular users is almost the same for both the spectrum access policies.
\end{itemize}

The rest of this paper is organized as follows: The system model is described in Section \ref{sec:sysModel}. In Section \ref{sec:access_prob}, the transmission probability of a D2D transmitter (i.e., the probability that the transmitter is able to harvest enough energy for channel inversion toward its intended receiver and the designated channel for transmission is available) is derived for different spectrum access policies for cellular communication. Section \ref{sec:out_prob} presents the analysis of the outage probability for D2D users and cellular users. Finally, the numerical results are presented in Section \ref{sec:results} before the paper is concluded in Section~\ref{sec:conc}. 

\begin{table}[!t]
\centering
\caption{List of Key Notations}
\label{notation}
\begin{tabular}{| l | l |}
\hline
\textbf{Notation} & \textbf{Definition}\\
\hline
\hline
$\mathbf{\Phi}_B$ & Point process of BSs\\
\hline
$\lambda_B$ & Spatial density of BSs\\
\hline
$P_B$ & Transmit power of BSs\\
\hline
$\rho_b$ & Receiver sensitivity of BS\\
\hline
$\mathbf{\Phi}_U$ & Point process of cellular users\\
\hline
$\lambda_U$ & Spatial density of cellular users\\
\hline
$N_u$ & Number of users per BS\\
\hline
$P_u$ & Transmit power of a cellular user\\
\hline
$\mathbf{\Phi}_D$ & Point process of D2D transmitters\\
\hline
$\lambda_D$ & Spatial density of D2D transmitters\\
\hline
$P_D$ & Transmit power of a D2D transmitter\\
\hline
$P_{\rm H}$ & Power harvested by a D2D transmitter\\
\hline
$a$ & RF-to-DC power conversion efficiency\\
\hline
$\rho_d$ & Receiver sensitivity of D2D receivers\\
\hline
$d_o$ & Max. inter-D2D users distance\\
\hline
$\mathcal{C}$ & Set of available channels\\
\hline
$\mathcal{C}_D$ & Set of available downlink channels\\
\hline
$\mathcal{C}_U$ & Set of available uplink channels\\
\hline
$c_d$ & D2D channel\\
\hline
$\alpha$ & Path-loss exponent of cellular links\\
\hline
$\beta$ & Path-loss exponent of D2D links\\
\hline
$h$ & Small-scale fading channel power gain\\
\hline
$\sigma_z^2$ & Noise power\\
\hline
$\gamma$ & Spectrum sensing threshold\\
\hline
$\mathcal{R}_{y}$ & Protection region\\
\hline
$r_P$ & Radius of protection region\\
\hline
$q_f$ & Prob. that a cellular user is assigned a channel\\
\hline
$q_c$ & Prob. that a BS uses channel $c_i\in \mathcal{C}\setminus c_d$\\
\hline
$q_d$ & Prob. that a BS uses channel $c_d$\\
\hline
$p_t$ & Transmission prob. for D2D transmitters\\
\hline
$p_f$ & Free-channel prob. for D2D transmitters\\
\hline
$p_s$ & Sufficient energy prob. for D2D transmitters\\
\hline
$\mathsf{O}_D$ & $\mathsf{SINR}$ outage prob. for D2D transmitters\\
\hline
$\mathsf{O}_D^{\rm tot}$ & Overall outage prob. for D2D transmitters\\
\hline
$\mathsf{O}_B$ & $\mathsf{SINR}$ outage prob. for cellular network\\
\hline
$\mathsf{O}_B^{\rm tot}$ & Overall outage prob. for cellular network\\
\hline
\end{tabular}
\end{table}

\section{System Model, Assumptions, and Methodology of Analysis}
\label{sec:sysModel}

The key mathematical notations used for the system model and analysis are summarized in Table \ref{notation}.

\subsection{Network Model}

We consider a cellular network in which macro BSs are overlaid with randomly located cognitive  D2D transmitters. The locations of the macro BSs are modeled by a homogeneous PPP $\mathbf{\Phi}_B = \{ x_i: i= 1, 2, \dots \}$ of spatial density $\lambda_B$ where $x_i \in \mathbb{R}^2$ denotes the location of the $i^{th}$ BS. All BSs are from the same type, i.e., macro BSs in this case, and transmit in the downlink with the same power level $P_B$. Cellular users are spatially distributed in $\mathbb{R}^2$ according to an independent homogeneous PPP $\mathbf{\Phi}_U = \{ u_i: i= 1, 2, \dots \}$ of density $\lambda_U$. Each cellular user associates with the closest BS, i.e., the BS from which she receives the strongest average signal. Cognitive D2D transmitters are also modeled by an independent two-dimensional PPP $\mathbf{\Phi}_D = \{ y_i: i= 1, 2, \dots \}$ with density $\lambda_D$. A D2D communication link is established only when the intended receiver is within a disc of radius $d_o$ and centered around the D2D transmitter\footnote{Since the problem of mode selection  is not within the scope of this work, we assume that each D2D transmitter has an intended receiver within a distance $d_o$ with probability $1$. For the mode selection in D2D Poisson networks,  refer to \cite{andrewsD2D,sawyD2D}.}. For the reliability of communication links, all users (i.e., D2D transmitters and cellular users) use channel inversion power control to adjust their transmit power by inverting the path-loss to insure that the average received signal power at the intended receiver (i.e., D2D receivers and BSs) is equal to its sensitivity. Here, we use $P_U$ and $P_D$ to denote the transmit power of an uplink cellular user and a D2D transmitter, respectively. It is assumed that all the D2D receivers have the same sensitivity $\rho_d$ and all the BSs have a sensitivity of $\rho_b$.  Saturation condition is assumed where each transmitter (i.e., BS in downlink, cellular user in uplink, or D2D transmitter) has at least one packet ready for transmission at the beginning of each time slot in a time-slotted transmission scenario. 

\subsection{Channel Model}
\begin{figure}[!t]
\centering
\includegraphics[width=7.2cm]{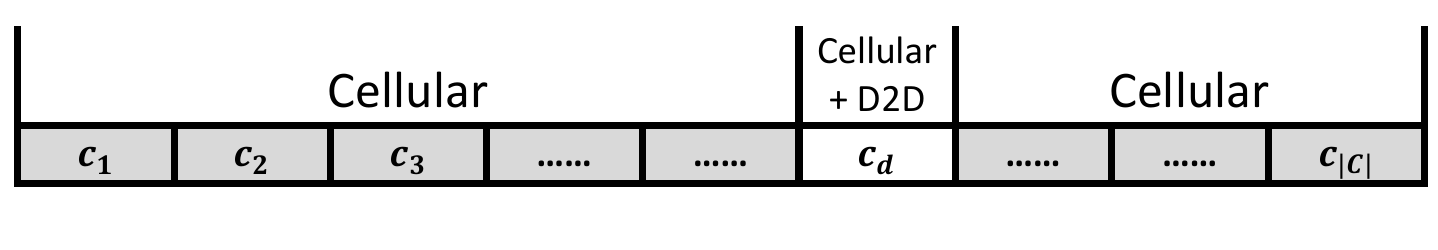}
\caption{Spectrum allocation for cellular and D2D transmissions.}
\vspace{-3mm}
\label{chs}
\end{figure}

The total available bandwidth is divided into a set of orthogonal channels $\mathcal{C} = \{ c_1, c_2, \dots, c_{|\mathcal{C}|} \}$ where $|\cdot|$ denotes the set cardinality. Furthermore, the set of channels $\mathcal{C}$ is partitioned into two disjoint subsets of channels $\mathcal{C}_D$ and $\mathcal{C}_U$ for downlink and uplink transmissions, respectively. While a cellular user can be served (in downlink or uplink) over any channel $c_i \in \mathcal{C}$ depending on the channel availability at the serving BS, all D2D transmissions take place on the same channel $c_d \in \mathcal{C}$, cf. Fig. \ref{chs}. Note that channel $c_d$ is not exclusive for D2D transmissions and can be used for cellular communication depending on the adopted spectrum access policy (i.e., RSA or PSA). Note also that $c_d$ can be either an uplink channel or a downlink channel used for cellular communication. In this work, we consider both cases when the D2D transmissions can take place either in one of the uplink channels or one of the downlink channels. At any BS, each associated cellular user is served by only one channel at most. In addition, there is no intra-cell interference assuming that each BS serves no more than one user in each channel. 

We assume that the power of the signal transmitted by a BS or a cellular user decays at a rate of $r^{-\alpha}$ where $\alpha$ is the path-loss exponent and $r$ is the propagation distance. On the other hand, the decay of the power transmitted by a D2D transmitter occurs at a rate of $r^{-\beta}$ where $\beta$ is the path-loss exponent\footnote{The channel model can be extended by incorporating multi-slope path-loss models as in \cite{andrewsMultiSlope}.}. Rayleigh fading is used to model small-scale fading over each channel where independence between channels is assumed. Hence, under the assumption that the channel gains are i.i.d., the power gain from a transmitter (i.e., BS, cellular user, or D2D transmitter) located at $x$ toward a generic point at $y$ on a channel $c_i$ is denoted by $h_x \sim \Exp(1)$. 

\subsection{Energy Harvesting Model}
All D2D transmitters are powered by  energy harvested from the ambient interference caused by the simultaneous cellular transmissions in the network (i.e., downlink and uplink transmissions). It is assumed that each D2D transmitter is equipped with an energy harvesting circuit that harvests RF power from all channels including both downlink and uplink channels. Therefore, the total power available for harvesting by a D2D transmitter located at a generic location $y \in \mathbb{R}^2$ can be expressed as
\begin{align}
P_{\rm H}(y) &= a \sum_{c \in \mathcal{C}_D} \sum_{x_i \in \mathbf{\tilde{\Phi}}_B(c)} P_B h_{x_i} {\|x_i - y\|}^{-\alpha} \nonumber\\
						&+ a \sum_{c \in \mathcal{C}_U} \sum_{u_i \in \mathbf{\tilde{\Phi}}_U(c)} P_u h_{u_i} {\|u_i - y\|}^{-\alpha} \label{eq:P_h}
\end{align}
where the first term represents the amount of RF power harvested from the concurrent downlink cellular transmissions and the second term is for the RF power harvested from the concurrent uplink cellular transmissions. It is worth mentioning that the amount of power received from other concurrent D2D transmissions at the harvesting unit is not considered in $P_{\rm H}$. In (\ref{eq:P_h}), $\mathbf{\tilde{\Phi}}_B(c)$ is a PPP with intensity $q_c \lambda_B$ that represents the set of BSs using channel $c \in \mathcal{C}_D$ where $q_c$ is the probability that a BS uses this channel\footnote{$q_c$, the spectrum access probability of a BS, will  be discussed in detail later in Section \ref{sec:access_prob}.}. $\mathbf{\tilde{\Phi}}_U(c)$ is a point process with intensity $q_c \lambda_B$ that represents the set of users using channel $c \in \mathcal{C}_U$. Note that, unlike $\mathbf{\tilde{\Phi}}_B(c)$, $\mathbf{\tilde{\Phi}}_U(c)$ is not a PPP due to the correlation among uplink cellular users. $0< a \leq 1$ is the efficiency of the conversion from RF to DC power, and $\| \cdot \|$ denotes the Euclidean distance.

 It is worth noting that a D2D transmitter may not harvest enough energy in one time slot to transmit with sufficient power since the power available for harvesting varies depending on the location of the D2D transmitter and the network statistics such as the channel power gains. Therefore, with a time-slotted ``harvest-then-transmit" strategy,  it is assumed that the RF energy harvesting and DC conversion circuits of a D2D transmitter are activated only when the available power in the time-slot is at least equal to the amount of power needed to invert the channel to its intended receiver. There is no energy storage assumed where a D2D transmitter can save the extra harvested energy for the next time slot.

\subsection{Model for Spectrum Sensing and  Transmission by D2D Users}

All D2D transmitters are assumed to be cognitive where each transmitter senses the state of the channel $c_d$ at the beginning of each time slot before it decides whether or not to use the channel for transmission. In this work, the main purpose of spectrum sensing is to avoid the interference that results from the nearby cellular transmissions on channel $c_d$. That is, the D2D transmitter does not use the channel $c_d$ if the received interference from any neighboring transmitter (i.e., BS in the downlink or cellular user in the uplink) on this channel is higher than a predefined sensing threshold $\gamma$; otherwise, the channel is available to be used by the D2D transmitter. Note that increasing the sensing threshold increases the probability to access the channel $c_d$ while increasing the aggregate interference at the same time. On the other hand, decreasing the sensing threshold provides more protection to the D2D transmission by decreasing the aggregate interference; however, it reduces the chance to access the channel $c_d$.  In other words, cognition provides a protection region around each D2D transmitter in which that D2D transmitter cannot use the channel $c_d$ if there is at least one active transmitter using this channel inside this region. 

In general, the protection region $\mathcal{R}_{y}(\gamma) \subset \mathbb{R}^2$ around a generic D2D transmitter located at $y$ follows a random shape. For example, if the D2D transmission takes place in  a downlink channel $c_d$, the protection region can be defined as
\begin{align}
\mathcal{R}_{y}(\gamma) &= \left\{ x \in \mathbb{R}^2 : P_B h_x {\| x - y \|}^{-\alpha} > \gamma \right\}  \nonumber\\
&= \left\{ x \in \mathbb{R}^2: \| x - y \| < r_{P}, r_{P} = \left(\frac{P_B h_x}{\gamma}\right)^{\frac{1}{\alpha}} \right\} \label{eq:cog_area_DL}
\end{align}
where $r_{P}$ represents the random radius of the protection region. On the other hand, if the D2D transmission takes place in an uplink channel $c_d$, the radius of the protection region is defined as
\begin{align}
r_{P} = \left(\frac{ P_u  h_u}{\gamma}\right)^{\frac{1}{\alpha}}  \label{eq:cog_area_UL}
\end{align}
where $P_u$ is the transmit power of the cellular user. Note that, unlike $P_B$, $P_u$ is random due to the uplink power control scheme used by the cellular users.
 
On average, the protection region can be approximated by a disc $\mathcal{B}_{y}(\bar{r}_{P}) \subset \mathbb{R}^2$ centered around the D2D transmitter with a radius $\bar{r}_{P} = \mathbb{E}_{P,h} \left[ r_P \right]$. That is, in the downlink scenario, the radius is defined such that
\begin{align}
\bar{r}_{P} &= \left(\frac{P_B}{\gamma}\right)^{\frac{1}{\alpha}} \Gamma\left(\frac{\alpha+1}{\alpha}\right), \quad \text{for $c_d \in \mathcal{C}_D$} \label{eq:r_P_DL}
\end{align}
where $\mathbb{E}_h \left[ h_x^m \right]	= \Gamma\left(1+m\right)$ for any $m\in \mathbb{R}^+$ is used to derive (\ref{eq:r_P_DL}) and $\Gamma(z) = \int_0^\infty t^{z-1} e^{-t} \text{d}t$ is the gamma function. Similarly, in the uplink scenario, $\bar{r}_{P}$ can be expressed as
\begin{align}
\bar{r}_{P} &= \mathbb{E}_P \left[ \left(\frac{P_u}{\gamma}\right)^{\frac{1}{\alpha}} \right] \Gamma\left(\frac{\alpha+1}{\alpha}\right) \nonumber\\
&= \frac{1}{2 \sqrt{\lambda_B}} \left(\frac{\rho_b}{\gamma}\right)^{\frac{1}{\alpha}} \Gamma\left(\frac{\alpha+1}{\alpha}\right), \quad \text{for $c_d \in \mathcal{C}_U$} \label{eq:r_P_UL}
\end{align}
where (\ref{eq:r_P_UL}) follows by using the $\frac{1}{\alpha}$-th moment of the transmit power of a cellular user, i.e.,  $\mathbb{E} \left[ P_u^\frac{1}{\alpha} \right] = \frac{\rho_b^\frac{1}{\alpha}}{2 \sqrt{\lambda_B}}$ when using channel inversion uplink power control \cite{sakrup}. 

\begin{figure}[!t]
\centering
\includegraphics[width=7cm]{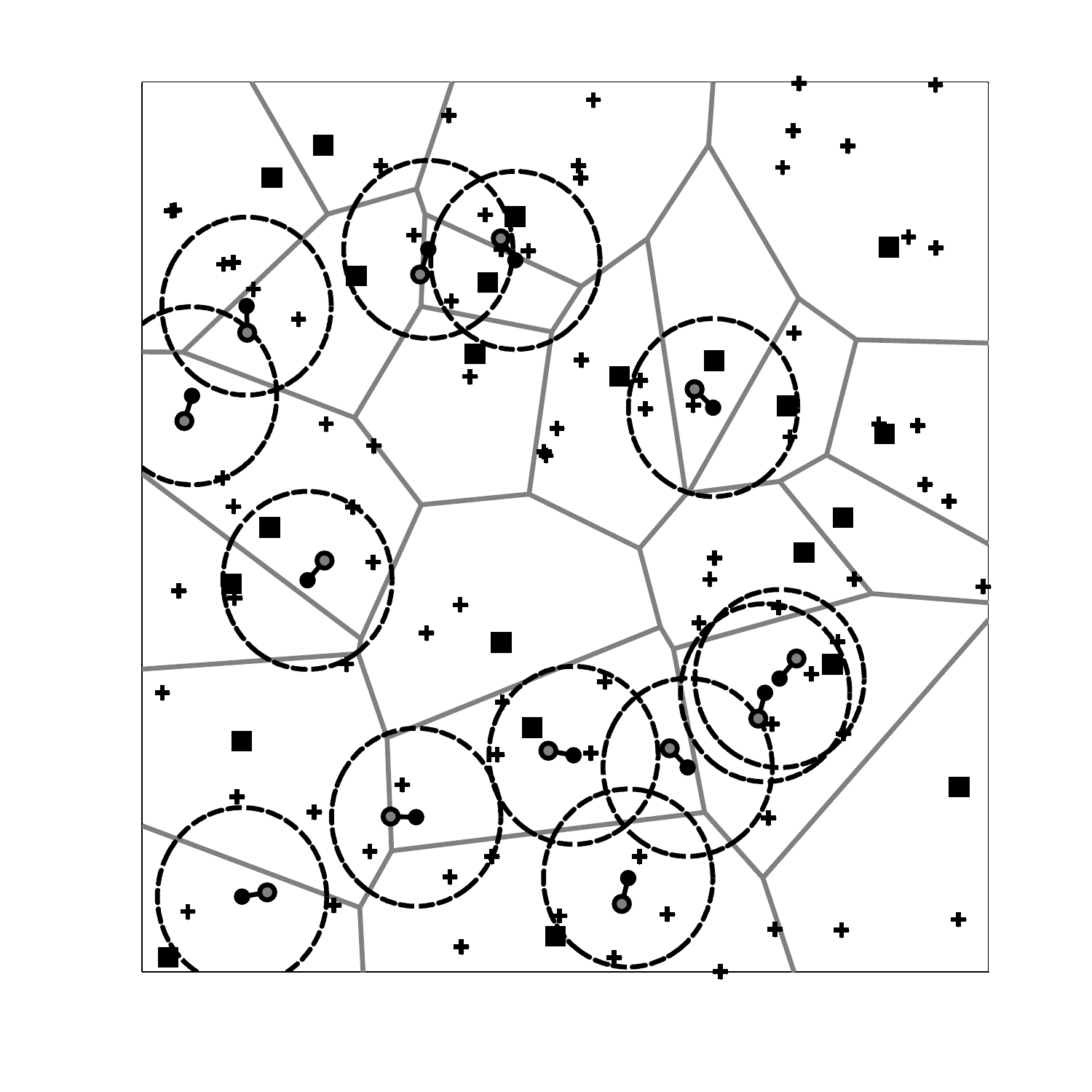}
\vspace{-3mm}
\caption{A realization of the described network model where squares represent the BSs, crosses represent the cellular users, and black lines represent the potential D2D links between the D2D transmitters (black dots) and D2D receivers (gray dots). Each D2D transmitter is surrounded by a protection region $\mathcal{B}(\bar{r}_{P})$ (dashed circles) with radius $\bar{r}_P$. The intensities are $\lambda_U = 4 \lambda_B$ and $\lambda_D = \lambda_B$.}
\vspace{-3mm}
\label{sysMod}
\end{figure}

Without loss of generality, Fig. \ref{sysMod} shows a realization of a cellular network whose macro BSs are underlaid with D2D transmitters. Note that not every D2D transmitter in Fig. \ref{sysMod} can use the D2D channel since it depends on the  allocation of channels to the cellular transmitters, which are located in the protection region of the D2D transmitter.

\subsection{Spectrum Access Model for Cellular Transmissions}
\label{subsec:access}
We consider two spectrum access policies, namely, random spectrum access (RSA) and prioritized spectrum access (PSA) policies, that define how the spectrum is assigned  for downlink and uplink transmissions to serve the cellular users. In RSA, in each cell,   any channel $c_i \in \mathcal{C}$ (including channel $c_d$ which is used for D2D transmission) can be independently and randomly assigned with the same probability to serve one of the cellular users. On the other hand, in PSA,  any channel $c_i \in \mathcal{C}\setminus \{c_d\}$ can be independently and randomly assigned to a cellular user as long as the number of  cellular users is less than the number of available channels $|\mathcal{C}|$. When the number of cellular users is higher than $|\mathcal{C}|-1$, only then,  $c_d$ will be assigned to a  cellular user.

\subsection{Methodology of Analysis}

Based on the system model described above, we aim at quantifying the performance of both D2D and cellular communications in terms of transmission probability for D2D users (i.e., the probability that there is a channel available for transmission and the amount of harvested energy is sufficient for D2D transmission) and $\mathsf{SINR}$ outage probability of both D2D and cellular users. The performance metrics are obtained for a test user located at the origin $(0,0)\in \mathbb{R}^2$; therefore, we drop the notation for the test user's location. According to Slivnyak's theorem~\cite{baccelli2010stochastic}, these results should be then valid for any generic user. We derive the spectrum access probabilities for the cellular transmissions and a generic D2D transmitter for both the RSA and PSA policies. Then, the probability density function ({\em pdf}) of  the interference power available for energy harvesting is derived to evaluate the probability that a D2D transmitter can harvest sufficient energy for transmission. Based on the obtained probabilities, the $\mathsf{SINR}$ outage probability is quantified for both the cellular  and D2D users. 

Note that there is no restriction on any BS to independently adopt either of the spectrum access policies. The intensity of BSs using each policy can be easily obtained by using the thinning property of a PPP; however, we consider the case when all BSs adopt the same spectrum access policy. 

\section{Spectrum Access Probabilities for Cellular Communication and Transmission Probabilities for D2D Users}
\label{sec:access_prob}

\subsection{Spectrum Access Probabilities for Cellular Communication}

For a cellular user, a connection (i.e., downlink or uplink connection) can only be established  when a channel $c_i\in \mathcal{C}$ is available. Therefore, we define $q_f$ as the probability that a BS has a free channel to assign to one of its associated users for uplink or downlink transmission.

Based on the spectrum access policies defined in Section \ref{subsec:access}, we derive the availability of any channel $c_i\in \mathcal{C}$ for cellular communication. Firstly, we denote the number of users associated to a BS as $N_u$ where the association policy is based on the nearest BS. According to \cite[Appendix A]{sawy2013mc}, the probability mass function ({\em pmf}) of the number of users served by a generic BS is obtained as
\begin{align}
\mathbb{P} \{ N_u =n \} = \zeta~ \frac{\Gamma(n+b)}{\Gamma(n+1) } \frac{\left(\mathbb{E}[N_u]\right)^n}{ \left(b + \mathbb{E}[N_u]\right)^{n+b}} \label{eq:pmf}
\end{align} 
where $\zeta = \frac{b^b}{\Gamma(b) }$ is a constant and $\mathbb{E}[N_u] = \frac{\lambda_U}{\lambda_B}$ is the average number of cellular users per BS. This expression is derived by approximating the area of a Voronoi cell by a gamma-distributed random variable with a shape parameter $b = 3.575$ and a scale parameter $\frac{1}{b \lambda_B}$.  Note that this expression is valid only when the cellular users are assumed to be spatially distributed according to an independent PPP and their associations to the BSs are based on the maximum average received signal power, i.e., each user associates with her nearest BS.

Note that all BSs in the cellular network share the same set of channels $\mathcal{C}$. While the number of channels assigned by each BS depends only on the number of its associated users $N_u$, the subset of channels used by each BS varies according to the adopted spectrum access policy. That is, the number of channels used by a BS for cellular communication is $\min\{ N_u, |\mathcal{C}| \}$ and the probability that there is a free channel to serve a cellular user does not depend on the adopted spectrum access policy. Thus, we use (\ref{eq:pmf}) to derive the probability $q_f$ that a cellular user finds a free channel available when it associates with a generic BS.

\begin{lem}\normalfont 
The probability that a cellular user is assigned a channel by her serving BS is 
\begin{align}
q_f &= 1  - \sum_{n=|\mathcal{C}|+1}^{\infty}  \frac{n - |\mathcal{C}|}{n } \mathbb{P} \{ N_u = n \} \label{eq:q_f}
\end{align}
where $\mathbb{P} \{ N_u = n \}$ is given by (\ref{eq:pmf}).
\label{lem_q_f}
\end{lem}
\begin{IEEEproof}
See \textbf{Appendix \ref{lem_q_f_proof}}.
\end{IEEEproof}

\vspace{0.2cm}
Intuitively, (\ref{eq:q_f}) shows that the probability that a BS is able to serve more cellular users increases with increasing number of channels $|\mathcal{C}|$, increasing intensity of BSs $\lambda_B$, or decreasing  intensity of cellular users $\lambda_U$. 

Note that the expression in (\ref{eq:q_f}) can be used for both  the RSA and PSA policies. To illustrate the impact of the adopted spectrum policy on the subset of the channels used for cellular communication, we define $q_c$ as the probability of a generic BS to assign a specific channel $c_i \in \mathcal{C}\setminus \{c_d\}$ to serve one of its associated users. We also define $q_d$ as the probability that a generic BS assigns the D2D channel $c_d$ to serve one of its associated users.

\vspace{0.2cm}
\subsubsection{Spectrum access probabilities for the RSA policy}
Since any BS randomly and independently assigns any channel $c_i \in \mathcal{C}$ with the same probability, the spectrum access probabilities $q_c^{\text{RSA}}$ and $q_d^{\text{RSA}}$ can be expressed as in the following lemma.

\begin{lem}\normalfont 
For the RSA policy, the probability that a BS uses a generic channel $c_i\in \mathcal{C}$ to serve one of its associated cellular users is given by
\begin{align}
q_c^{\text{RSA}} = 1 - \sum_{n=0}^{|\mathcal{C}|-1} \frac{|\mathcal{C}|-n}{|\mathcal{C}|}  \mathbb{P} \{ N_u = n \} \label{eq:q_c_RSA}
\end{align}
where $q_d^{\text{RSA}} = q_c^{\text{RSA}}$ and $\mathbb{P} \{ N_u = n \}$ is given by (\ref{eq:pmf}).
\label{lem_RSA}
\end{lem}
\begin{IEEEproof}
See \textbf{Appendix \ref{lem_RSA_proof}}.
\end{IEEEproof}

\vspace{0.2cm}
\subsubsection{Spectrum access probabilities for the PSA policy}
With PSA, since a BS  uses $c_d$ only when it runs out of the rest of the available channels, the probability of using a channel $c_i\in \mathcal{C} \setminus \{c_d\}$ by a generic BS is different from that of using $c_d$, i.e., $q_c^{\text{PSA}} \neq q_d^{\text{PSA}}$. The expressions for $q_c^{\text{PSA}}$ and $q_d^{\text{PSA}}$ are presented in the following lemma.

\begin{lem}\normalfont 
For the PSA policy, the probability that a BS uses a generic channel $c_i\in \mathcal{C} \setminus \{c_d\}$ to serve one of its associated cellular users is given by
\begin{align}
q_c^{\text{PSA}} = 1 - \sum_{n=0}^{|\mathcal{C}|-1} \frac{|\mathcal{C}|-n-1}{|\mathcal{C}|-1}  \mathbb{P} \{ N_u = n \} \label{eq:q_c_PSA} 
\end{align}
while the probability that a BS has to use $c_d$ to serve one of its associated cellular users is given by
\begin{align}
q_d^{\text{PSA}} = 1 - \sum_{n=0}^{|\mathcal{C}| - 1} \mathbb{P} \{ N_u =n \} \label{eq:q_d_PSA}
\end{align}
where $\mathbb{P} \{ N_u = n \}$ is given by (\ref{eq:pmf}).
\label{lem_PSA}
\end{lem}
\begin{IEEEproof}
See \textbf{Appendix \ref{lem_PSA_proof}}.
\end{IEEEproof}

\vspace{0.2cm}
By comparing the results presented in \textbf{Lemma \ref{lem_RSA}} and \textbf{Lemma \ref{lem_PSA}}, it can be noticed that, when compared to RSA, PSA lowers the probability of assigning the D2D channel $c_d$ for cellular communication. The decrease in the access probability of $c_d$ is
\begin{align}
q_d^{\text{RSA}} - q_d^{\text{PSA}} &= \sum_{n=0}^{|\mathcal{C}| - 1} \frac{n}{|\mathcal{C}|}\mathbb{P} \{ N_u =n \}. \label{eq:q_d_reduction}
\end{align}
This reduction in the access probability of $c_d$ comes at the expense of congesting other channels in $\mathcal{C} \setminus \{c_d\}$. The increase in the access probability of a channel $c_i\in \mathcal{C} \setminus \{c_d\}$ is
\begin{align}
q_c^{\text{PSA}} - q_c^{\text{RSA}} &= \frac{q_d^{\text{RSA}} - q_d^{\text{PSA}}}{|\mathcal{C}|-1}. \label{eq:q_c_increase}
\end{align}

\subsection{Transmission Probability for D2D Transmitters}

As described above, a D2D transmitter can access the channel $c_d$ only when it has not been assigned to any user  by any of the BSs in the protection region defined in (\ref{eq:cog_area_DL}) (when $c_d$ is a downlink channel) or (\ref{eq:cog_area_UL}) (when $c_d$ is an uplink channel). In other words, the probability that a D2D transmitter is able to use channel $c_d$ is equal to the probability that the number of BSs, which use this channel in the protection region of this D2D transmitter, is zero. Let us denote this probability by $p_f$. The D2D transmitter also needs to harvest sufficient energy to perform channel inversion, the probability of which is denoted by $p_s$. Hence, the transmission probability of a D2D transmitter can be defined as
\begin{align}
p_t = p_s p_f. 
\label{eq:p_t}
\end{align} 

\vspace{0.2cm}
\subsubsection{Calculation of $p_f$}
The expression for $p_f$  is provided in the following lemma.

\begin{lem}\normalfont 
For a generic D2D transmitter, the probability that the D2D channel (i.e., $c_d$) is free is given by
\begin{align}
p_f = \exp [{-\theta q_d}] \label{eq:p_f}
\end{align}
where 
\begin{align}
\theta = \left\{
\begin{array}{l l}
	\pi \lambda_B  \left(\frac{P_B}{ \gamma}\right)^{\frac{2}{\alpha}} \Gamma\left(\frac{\alpha+2}{\alpha}\right), \quad &\text{for $c_d \in \mathcal{C}_D$}\\	
	\left(\frac{\rho_b}{ \gamma}\right)^{\frac{2}{\alpha}} \Gamma\left(\frac{\alpha+2}{\alpha}\right), \quad &\text{~for $c_d \in \mathcal{C}_U$}.
\end{array}\right. \label{eq:theta}
\end{align}
Here, $q_d$ is the probability that the D2D channel is used by a generic BS for cellular communication, which is given by (\ref{eq:q_c_RSA}) and (\ref{eq:q_d_PSA}) for the RSA and PSA policies, respectively.
\label{lem_p_f}
\end{lem}
\begin{IEEEproof}
See \textbf{Appendix \ref{lem_p_f_proof}}.
\end{IEEEproof}

\vspace{0.2cm}
\subsubsection{Calculation of $p_s$}
Although each D2D transmitter has an intended receiver within a circle with a radius $d_o$, we consider the worst-case scenario where the receiver is at the boundary of the circle (i.e., at a distance $d_o$). Note that this assumption provides an upper bound on the amount of transmit power required to perform channel inversion uplink power control, hence, it provides a lower bound on the probability of harvesting sufficient energy. Relaxing this assumption complicates the derived expressions without adding more insights \cite{weber}.

For a D2D transmitter with an intended receiver at a distance $d_o$, the minimum required transmit power that results in a received signal of $\rho_d$ at the intended receiver by channel inversion can be obtained as: $P_D =  \rho_d d_o^\beta$. We define the probability $p_s$ that a D2D transmitter harvests sufficient energy to perform channel inversion as follows:
\begin{align}
p_s = \mathbb{P} \left[ P_{\rm H} > P_D \right], \label{eq:p_s_def}
\end{align}
where $P_{\rm H}$ is defined in (\ref{eq:P_h}). The following lemma provides an expression for $p_s$ for a general path-loss exponent $\alpha$.

\begin{lem}\normalfont 
The probability that a typical D2D transmitter harvests sufficient energy for transmission is 
\begin{align}
p_s = \frac{\alpha}{2\pi} \int_0^\infty &\frac{1}{u} \exp\left[- \rho_d d_o^\beta \left(\frac{u}{\kappa_3 }\right)^{\frac{\alpha}{2}}\right] \times  \label{eq:p_s}\\
&\exp\left[- u \cos\left(\frac{2\pi}{\alpha}\right)\right] \sin\left(u \sin\left(\frac{2\pi}{\alpha}\right)\right) \text{d}u \nonumber
\end{align}
where $\kappa_3 = \kappa_1+\kappa_2$ and
\begin{align}
\kappa_1 &= \frac{2 \pi q_c^{\rm D} \lambda_B (a P_B)^{\frac{2}{\alpha}}\Gamma\left(\frac{\alpha-2}{\alpha} \right)}{\alpha} \sum_{k=1}^{|\mathcal{C}_D|} \frac{\Gamma\left(k-\frac{\alpha-2}{\alpha}\right)}{(k-1)!},   \label{eq:kappa_1}
\end{align}
\begin{align}
\kappa_2 &= \frac{2\pi q_c^{\rm U} |\mathcal{C}_U| (a \rho_b)^{\frac{2}{\alpha}}}{ \alpha \sin\left(\frac{2\pi}{\alpha}\right)}.  \label{eq:kappa_2}
\end{align}
Here, $q_c^{\rm D}$ and $q_c^{\rm U}$ are the access probabilities corresponding to a channel $c\in \mathcal{C}_D$ and a channel $c\in \mathcal{C}_U$, respectively. They are obtained based on (\ref{eq:q_c_RSA}) by using $|\mathcal{C}_D|$ and $|\mathcal{C}_U|$, respectively.
\label{lem_p_s}
\end{lem}
\begin{IEEEproof}
See \textbf{Appendix \ref{lem_p_s_proof}}.
\end{IEEEproof}

\vspace{0.2cm}
Furthermore, in the following corollary we show that the probability of harvesting sufficient energy $p_s$ can be presented in a closed-form expression in the special case when $\alpha = 4$. 
\begin{col}\normalfont 
The probability that a typical D2D transmitter harvests sufficient energy for transmission when $\alpha=4$ is    
\begin{align}
p_s &= \erf \left(\frac{\kappa_3}{2 \sqrt{\rho_d d_o^\beta}} \right) \label{eq:p_s_4}
\end{align}
where $\erf(z) = \frac{2}{\sqrt{\pi}} \int_0^z e^{-t^2} \text{d}t$ is the error function and $\kappa_3$ can be reduced as 
\begin{align}
\kappa_3 = \frac{1}{2} \pi^2 q_c^{\rm D} \lambda_B \sqrt{a P_B} \sum_{k=0}^{|\mathcal{C}_D|-1} \frac{1}{4^k} \binom{2k}{k}  + \frac{1}{2}\pi q_c^{\rm U} |\mathcal{C}_U|  \sqrt{a \rho_b}.  \label{eq:kappa_3_}
\end{align}
\label{col_p_s_1}
\end{col}
\begin{IEEEproof}
When $\alpha = 4$, the {\em cdf} of aggregate received interference follows a L\'{e}vy distribution with a location parameter $0$ and a scaling parameter $\frac{\kappa_3^2}{2}$. Hence, $F_{P_{\rm H}}(x) = \erfc(\frac{\kappa_3}{2\sqrt{x}})$ where $\erfc(z) = 1 - \erf(z)$ is the complementary error function. 
\end{IEEEproof}

\textbf{Remark:} Note that: \textbf{1)} from (\ref{eq:p_s}) and (\ref{eq:p_s_4}), it can be seen that $p_s$ is an increasing function of $\kappa_i$ for $i=\{1,2,3\}$. \textbf{2)} all $q_c$s are decreasing functions of the number of channels $|\mathcal{C}|$. \textbf{3)} the summations in (\ref{eq:kappa_1}) and (\ref{eq:kappa_3_}) are increasing functions of the number of channels. Thus, it can be easily proven that $\kappa_i$s are discrete concave functions and there exists an optimal value of $|\mathcal{C}|$ that maximizes both $\kappa_3$ and $p_s$. Note also that, increasing the number of available channels to a very large value does not necessarily improve $p_s$ since not all channels will be used by each BS and $p_s$ becomes limited by the ratio of $\lambda_B$ and $\lambda_U$. This can be seen in (\ref{eq:kappa_2}) by noting that $\lim \limits_{|\mathcal{C}|\rightarrow \infty} q_c |\mathcal{C}| = \frac{\lambda_U}{\lambda_B}$.

\section{Analysis of $\mathsf{SINR}$ Outage Probability}
\label{sec:out_prob}

In this section, we characterize the $\mathsf{SINR}$ outage probability for both cellular and D2D receivers. A receiver is considered in outage if the $\mathsf{SINR}$ falls below a given threshold $\tau$. Note that $\tau$ can be chosen based on users'  QoS requirements.

\subsection{Outage Probability for a D2D Receiver}

For the analysis of $\mathsf{SINR}$ outage probability, we consider a typical D2D receiver at the origin $(0,0) \in \mathbb{R}^2$ while the results hold for any generic D2D receiver. In addition, the $\mathsf{SINR}$ outage probability is derived given that the corresponding D2D transmitter has sufficient energy to invert the channel and the D2D channel $c_d$ is not used within its protection region. Each D2D receiver suffers from two sources of interference, i.e., cellular network and D2D network. For the cellular network, the aggregate interference results from all macro BSs (if $c_d$ is a downlink channel) or all cellular users (if $c_d$ is an uplink channel) that use channel $c_d$. Hence, we define a homogenous PPP $\mathbf{\tilde{\Phi}}_B(c_d)$ with intensity $q_d \lambda_B$ that represents the set of BSs that use channel $c_d$ and another PPP $\mathbf{\tilde{\Phi}}_U(c_d)$ with intensity $q_d \lambda_B$ that represents the set of users who use channel $c_d$ where $q_d$ is defined in (\ref{eq:q_c_RSA}) and (\ref{eq:q_d_PSA}), respectively, for the RSA and PSA policies. For the D2D network, the aggregate interference results only from other D2D transmitters that have sufficient energy to invert the channel to their intended receivers and can transmit on $c_d$. Hence, the interfering D2D transmitters do not constitute a homogeneous point process anymore and analytical  characterization of  interference is not possible\footnote{This point process is called ``point hole process'' since the active D2D transmitters cannot be at a distance less than $r_p$ to a cellular transmitter that uses channel $c_d$. The analysis of such a point process is intractable due to unknown probability generating functional \cite{martin_int}.}. Therefore, for analytical tractability, we ignore the correlation among the locations of the interfering D2D transmitters and approximate the point process by a homogenous PPP $\mathbf{\tilde{\Phi}}_D$ with the same intensity $p_t \lambda_D$ where $p_t = p_s p_f$. This approximation will be validated in Section \ref{sec:results}.

For a typical D2D receiver, the $\mathsf{SINR}$ can be written as
\begin{align}
\mathsf{SINR}_D &= \frac{\rho_d h_{y_o}}{I_B + I_D + \sigma_z^2} \label{eq:SINR_D}
\end{align}
where $y_o$ is the corresponding D2D transmitter at a distance $d_o$, $h_{y_i}$ is the small-scale fading coefficient between the typical D2D receiver and its D2D transmitter, and $\sigma_z^2$ is the variance of the additive noise at the receiver where no specific noise distribution is assumed. $I_B$ and $I_D$ denote the aggregate interference  on $c_d$ resulting from the cellular network (downlink or uplink) and other D2D transmitters, respectively.

Using the instantaneous $\mathsf{SINR}$ in (\ref{eq:SINR_D}), we can obtain the $\mathsf{SINR}$ outage probability $\mathsf{O}_D$ for a typical D2D receiver for both cases when $c_d$ is either a downlink channel or an uplink channel. The outage probability is defined as the probability that the  $\mathsf{SINR}$ at the receiver is less than a predefined threshold $\tau$, i.e., $\mathsf{O}_D = \mathbb{P} \left[ \mathsf{SINR}_D \le \tau \right]$. The $\mathsf{SINR}$ outage probability is obtained in the following theorem. Note that the coverage is the complementary event of the outage event. 

\begin{thm}\normalfont 
The $\mathsf{SINR}$ outage probability for a typical D2D receiver is given by $\mathsf{O}_D = 1- \exp \left[- \mathcal{K}_1 \right]$, where
\begin{align}
\mathcal{K}_1	& = \frac{ \tau \sigma_z^2}{\rho_d}   + \frac{2 \pi^2 d_o^2 p_s p_f \lambda_D}{\beta \sin\left( \frac{2\pi}{\beta} \right)} \tau^{\frac{2}{\beta}} \\
&+ \frac{2\pi q_d \lambda_B}{\alpha-2} \mathcal{G}\left[\left(\frac{\rho_d}{\gamma \tau}\right)^{\frac{1}{\alpha}} \Gamma\left(\frac{\alpha+1}{\alpha}\right),\alpha\right] \left( \frac{\bar{P}}{\rho_d}\tau \right)^{\frac{2}{\alpha}} \nonumber
\end{align}
such that 
\begin{align}
\bar{P} &= \left\{
\begin{array}{l l}
P_B, & \text{for $c_d \in \mathcal{C}_D$}  \\
\frac{\rho_b}{(\pi \lambda_B)^{\frac{\alpha}{2}}}, & \text{for $c_d \in \mathcal{C}_U $}
\end{array}\right. \label{eq:P_bar}
\end{align}
where $\mathcal{G}[y,\alpha] = y^{2-\alpha} \,_2F_1\left[1,\frac{\alpha-2}{\alpha};\frac{2\alpha-2}{\alpha};-y^{-\alpha}\right]$ and $\,_2F_1\left[a,b;c;x\right]$ is Gauss Hypergeometric function.
\label{thm_O_D}
\end{thm}
\begin{IEEEproof}
See \textbf{Appendix \ref{thm_O_D_proof}}.
\end{IEEEproof}

Combining the results in \textbf{Theorem \ref{thm_O_D}} and \textbf{Lemma \ref{lem_p_f}}, we can obtain the reference value of $\lambda_B$ (as given in the lemma below) that can be used to decide whether to use a downlink channel or an uplink channel for D2D transmissions.
\begin{lem}\normalfont 
The spatial density of BSs beyond which using an uplink channel is more beneficial for cognitive D2D communication compared to using a downlink channel is given by
\begin{align}
\lambda_B^{\rm ref} &= \frac{1}{\pi} \left(\frac{\rho_b}{P_B}\right)^{\frac{2}{\alpha}}. \label{eq:lambda_ref}
\end{align}
\label{lem_lambda_ref}
\end{lem}
\begin{IEEEproof}
From (\ref{eq:theta}) and (\ref{eq:P_bar}), at $\lambda_B^{\rm ref}$, $\theta$ and $\bar{P}$ are the same for both downlink and uplink cases. Increasing $\lambda_B$ beyond this value increases $\theta$ for the downlink case and reduces $\mathcal{K}_1$ for the uplink case, and vice versa. This improves the overall performance of  D2D communication (i.e., higher $p_f$ and lower $\mathsf{SINR}$ outage) when $c_d$ is an uplink channel compared to the case when $c_d$ in a downlink channel.
\end{IEEEproof}

Now, we define the overall outage probability for a generic D2D receiver as
\begin{align}
\mathsf{O}_D^{\rm tot} &= 1 - p_t + p_t \mathsf{O}_D \label{eq:O_D_tot}
\end{align}
where this overall outage includes all the three possible events that cause outage, i.e., the event of not harvesting sufficient energy, the event of not finding the channel $c_d$ to be free, and the event of receiving an insufficient level of $\mathsf{SINR}$.

\subsection{Outage Probability for a Cellular User}

Since each cellular user associates with the closest BS, we can define the BS $x_o$ to which a cellular user located at $u_i$ is associated with as follows:
\begin{align}
x_o &= \arg \max_{x_i \in \mathbf{\Phi}_B} \left\{ \|x_i - u_i\|^{-\alpha} \right\}. \label{eq:xs}
\end{align}
Note that this association policy ensures that a generic cellular user is closer to its serving BS than any interferer (i.e., BS or cellular user). Hence, the received $\mathsf{SINR}$ at a typical cellular user (in downlink) or a macro BS (in uplink) on a generic channel $c \in \mathcal{C}$ can be written as
\begin{align}
\mathsf{SINR}_B(c) &= \left\{
\begin{array}{l l}
{\displaystyle \frac{P_B h_{x_o} \|x_o\|^{-\alpha}}{ I_B^{\rm D}(c) + I_D  \cdot \mathbf{1}_{c=c_d}  + \sigma_z^2}}, & \text{for $c \in \mathcal{C}_D$}  \\
{\displaystyle \frac{\rho_b h_{x_o}}{ I_B^{\rm U}(c) + I_D \cdot \mathbf{1}_{c=c_d} + \sigma_z^2}}, & \text{for $c \in \mathcal{C}_U $}
\end{array}\right.
\end{align}
where $x_o$ is the serving BS and $\mathbf{1}_{A}$ is the indicator function that equals $1$ only when $A$ is true and $0$ otherwise.
 
For the interference in the downlink network, $\tilde{\mathbf{\Phi}}_B(c)$ represents a point process of BSs using channel $c$, which has an intensity of $q_c \lambda_B$ for $c\in \mathcal{C}_D\setminus \{c_d\}$ and $q_d \lambda_B$ for $c = c_d$. For the uplink transmissions, $\tilde{\mathbf{\Phi}}_U(c)$ represents a point process of users using channel $c$, which has an intensity of $q_c \lambda_B$ for $c\in \mathcal{C}_U\setminus \{c_d\}$ and $q_d \lambda_B$ for $c = c_d$.  It is worth mentioning that the $\mathsf{SINR}$ outage probability $\mathsf{O}_B$ for a typical cellular user depends on the adopted spectrum access probability for cellular communication since the $\mathsf{SINR}$ depends on channel $c$. Using the instantaneous $\mathsf{SINR}$, we can obtain the $\mathsf{SINR}$ outage probability $\mathsf{O}_B$ for a typical cellular user on channel $c$ as
\begin{align}
\mathsf{O}_B(c) &= \mathbb{E}_x \left[ \mathbb{P} \left[ \mathsf{SINR}_B \le \tau \right]\right]. \label{eq:outage_gen_B}
\end{align}

The overall $\mathsf{SINR}$ outage probability is obtained in the following theorem.
\begin{thm}\normalfont 
For an interference-limited cellular network with $\alpha=\beta$, the $\mathsf{SINR}$ outage probability for a typical cellular user on channel $c$ is given by
\begin{align}
\mathsf{O}_B(c) &= \left\{
\begin{array}{l l}
{\displaystyle 1 - \frac{\pi \lambda_B}{\pi \lambda_B(1+\mathcal{K}_2) + \left(\frac{\rho_d}{P_B}\right)^{\frac{2}{\beta}}\mathcal{K}_3 }}, & \text{for $c \in \mathcal{C}_D$}  \\
{\displaystyle1 - \exp \left[-\left(\mathcal{K}_2 + \left(\frac{\rho_d}{\rho_b}\right)^{\frac{2}{\beta}} \mathcal{K}_3\right)  \right]}, & \text{for $c \in \mathcal{C}_U $}
\end{array}\right.
\label{eq:O_B}
\end{align}
where 
\begin{align}
\mathcal{K}_2 &= \frac{2 \hat{q} }{\alpha-2} \mathcal{G}\left[\left(\frac{1}{\tau}\right)^{\frac{1}{\alpha}},\alpha\right] \tau^{\frac{2}{\alpha}},\\
\mathcal{K}_3 &= \frac{2 \pi^2 p_s p_f \lambda_D d_o^2}{\beta \sin\left(\frac{2\pi}{\beta}\right)} \tau^{\frac{2}{\beta}} \cdot \mathbf{1}_{c=c_d},
\end{align}
and
\begin{align}
\hat{q} &= \left\{
\begin{array}{l l}
q_c, & \text{for } c\in \mathcal{C} \setminus \{c_d\} \\
q_d, & \text{for } c = c_d
\end{array}\right. \label{eq:q_hat}
\end{align}
in which $q_c$ and $q_d$ are given for both the RSA and PSA policies in (\ref{eq:q_c_RSA})-(\ref{eq:q_d_PSA}).
\label{thm_O_B}
\end{thm}
\begin{IEEEproof}
See \textbf{Appendix \ref{thm_O_B_proof}}.
\end{IEEEproof}

To average the $\mathsf{SINR}$ outage probability over all channels, by using the law of total probability, we can obtain the average $\mathsf{SINR}$ outage probability $\mathsf{O}_B^{\rm avg}$. Note that the probability of a user to be served on a certain channel can be obtained in the same manner as $q_c$ and $q_d$. Note also that, for RSA, the $\mathsf{SINR}$ outage probabilities corresponding to all channels are the same. To obtain the overall outage probability for a cellular user $\mathsf{O}_B^{\rm tot}$, we can incorporate the probability that a BS has at least one free channel for each of its associated users, i.e., $q_f$, as defined in (\ref{eq:q_f}).

\section{Numerical Results and Discussions}
\label{sec:results}

\subsection{System Parameters}
We use the obtained closed-form expressions to evaluate system performance in different scenarios for both the access policies (i.e., RSA and PSA) and both the cases when $c_d$ is a downlink channel or an uplink channel. Hence, we have four possible scenarios as follows: Downlink-RSA, Downlink-PSA, Uplink-RSA, and Uplink-PSA. The performance metrics include the probability of harvesting sufficient energy ($p_s$), channel access probability for a D2D user ($p_f$), transmission probability for a D2D user ($p_t$), $\mathsf{SINR}$ outage probability ($\mathsf{O}_D$), as well as the overall outage probability ($\mathsf{O}_D^{\rm tot}$) for D2D users. In addition, the channel access probability ($q_f$) and $\mathsf{SINR}$ outage probability are used to show the effect of the proposed spectrum policies on the performance of cellular users. Moreover, Monte Carlo simulations are used to validate the PPP assumptions used in deriving all the expressions.

For numerical evaluation, unless otherwise stated, the transmit power of a macro BS is assumed to be $37$ dBm while the thermal noise power $\sigma_z^2$ is $-104$ dBm. The receiver sensitivity of macro BSs $\rho_b$ is $-70$ dBm. The spatial densities of macro BSs, cellular users, and D2D transmitters are $\lambda_B = 1$ BS/km$^2$, $\lambda_U = 10 \lambda_B$ users/km$^2$, and $\lambda_D = 20$ users/km$^2$, respectively.  Independent and identically distributed Rayleigh fading with unit variance is considered for all links. The path-loss exponent for the cellular propagation is $\alpha = 4$ and that for  D2D transmission is $\beta = 3$. The total number of channels is $|\mathcal{C}| = 10$ channels. The reference distance $d_o$ is set to $10$ m and for the evaluation of outage probability, the threshold $\tau$ is set to $0$ dB. For Monte Carlo simulations, we choose a simulation area of $20$ km $\times 20$ km in order to guarantee a negligible matching error and boundary effects. The simulations are carried out by using MATLAB and the results are averaged over $10,\!000$ iterations. To validate our results, none of the PPP assumptions made during the analytical derivations are retained during the simulation where all user selection and scheduling are performed by the simulator. In the following figures, the simulation results are represented by curves with black cross markers, i.e., ``+''.

\subsection{Transmission Probability for a D2D Transmitter}

\begin{figure}[!t]
\centering
\includegraphics[width=8.7cm]{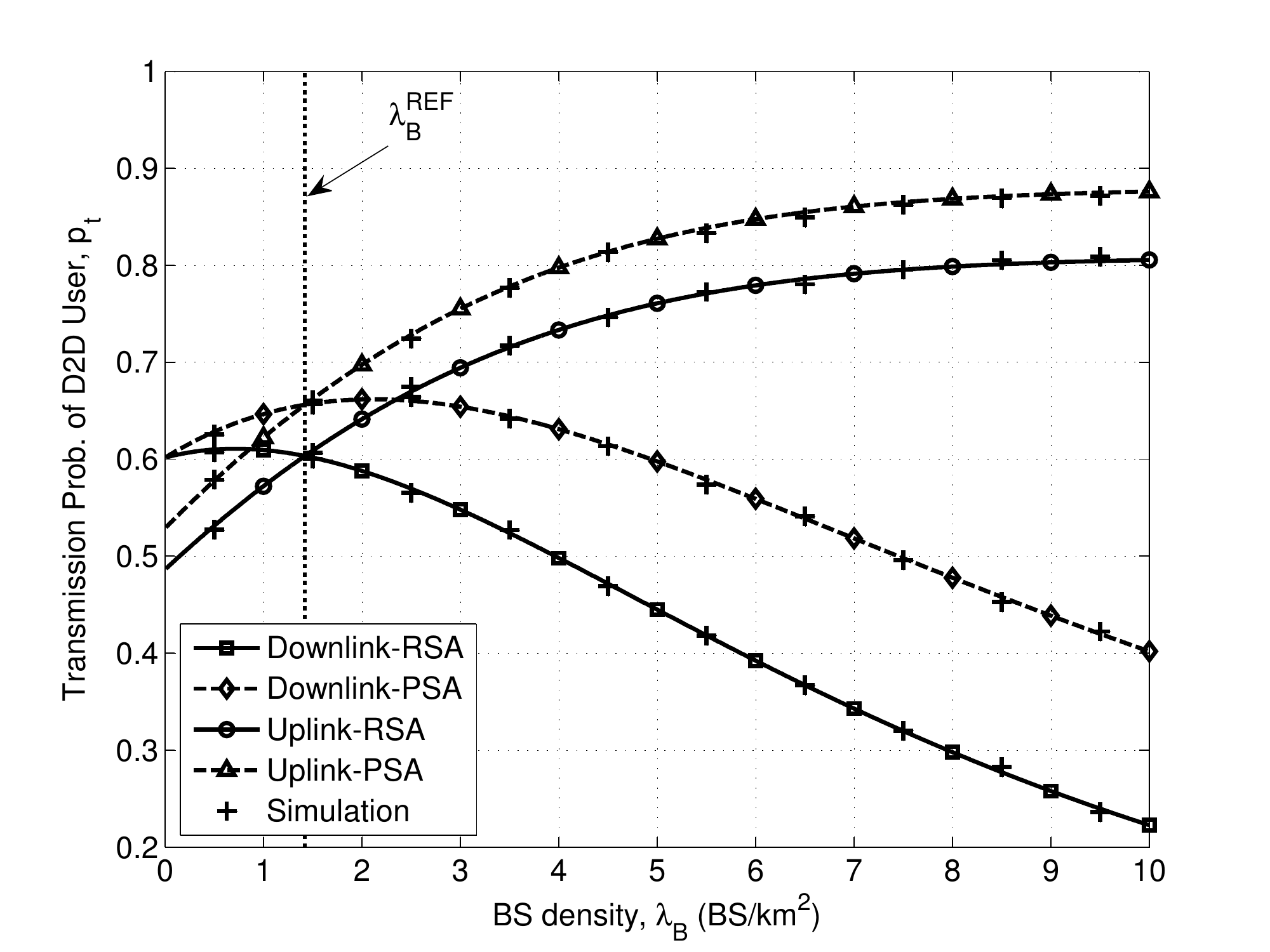}
\caption{The transmission probability $p_t$ for a generic D2D transmitter vs. the intensity of BSs when $c_d$ is a downlink or an uplink channel. The network parameters are $\rho_d = -80$ dBm, $d_o = 10$ m, $\gamma = -60$ dBm, and $\lambda_U = 10 \lambda_B$. The results are shown for both  the RSA and PSA policies.}
\label{p_f_vs_lambda_B_2}
\end{figure}

\begin{figure}[!t]
\centering
\includegraphics[width=8.7cm]{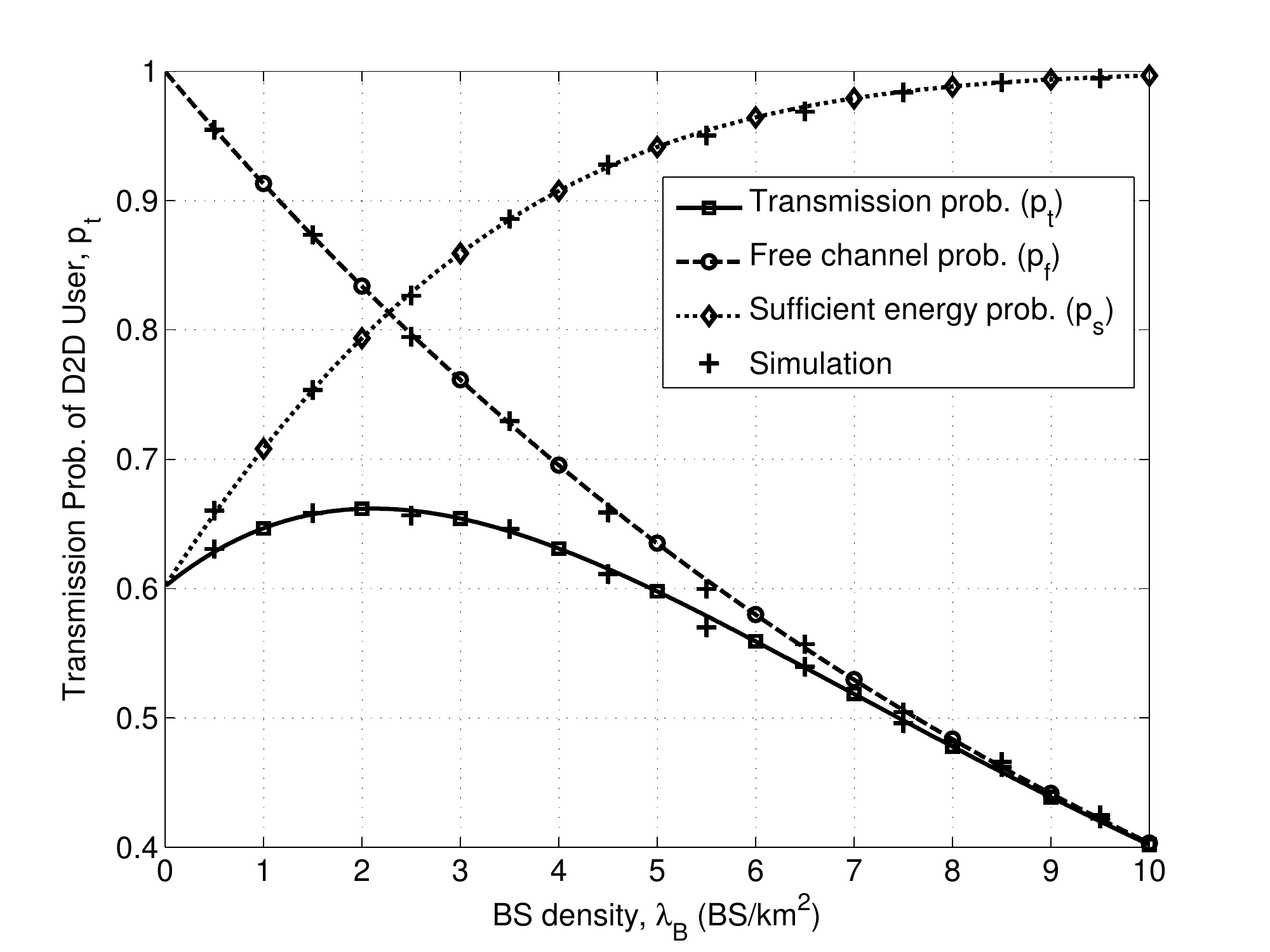}
\caption{The transmission probability $p_t$ for a generic D2D transmitter vs. the intensity of BSs when $c_d$ is a downlink channel and PSA policy is used. The network parameters are $\rho_d = -80$ dBm, $d_o = 10$ m, $\gamma = -60$ dBm, and $\lambda_U = 10 \lambda_B$.}
\label{p_f_vs_lambda_B_1}
\end{figure}

Fig. \ref{p_f_vs_lambda_B_2} shows the effect of varying the density of macro BSs on the transmission probability for the four different scenarios.  Fig. \ref{p_f_vs_lambda_B_1}  elaborates more on this effect  for one of these scenarios (i.e., Downlink-PSA) and the same explanations hold for the all other scenarios. From \textbf{Lemma \ref{lem_p_f}}, it can be seen that increasing $\lambda_B$ has two effects on the probability that the D2D channel is free ($p_f$). The first effect is captured in the term $\theta$ which reflects the increase in the number of BSs inside the protection region of the D2D transmitter. The second effect is the reduction in access probability of D2D channel $c_d$ by the macro BSs where this effect is included in the term $q_d$. Note that the first effect is independent of the spectrum access policy and depends on whether $c_d$ is a downlink or an uplink channel. On the other hand, $q_d$ is a function of the adopted spectrum access policy and independent of $c_d$. Regarding $q_d$, since the average number of cellular users per BS (i.e., $\mathbb{E}[N_u]=\frac{\lambda_U}{\lambda_B}$) is fixed, it becomes independent of $\lambda_B$ as can be noticed in (\ref{eq:q_c_RSA}), (\ref{eq:q_d_PSA}), and (\ref{eq:pmf}). On the other hand, since the transmit power of BSs is constant, $\theta$ becomes an increasing function of $\lambda_B$ when $c_d$ is a downlink channel. When $c_d$ is an uplink channel, since cellular users use channel inversion power control and the average transmit power is a decreasing function of $\lambda_B$, the overall effect makes $\theta$ independent of $\lambda_B$ in this case as can be seen in (\ref{eq:theta}). 

When $c_d$ is a downlink channel, it can be concluded that the effect of increasing $\lambda_B$ is always dominated by the increase in $\theta$ (as shown in Fig. \ref{p_f_vs_lambda_B_1} for Downlink-PSA). As a result, $p_f$ decreases. On the other hand, for the uplink case, it can be concluded that $p_f$ is constant since both $\theta$ and $q_d$ are independent of $\lambda_B$. Furthermore, from \textbf{Lemma \ref{lem_p_s}},  the probability that a D2D transmitter can harvest sufficient energy ($p_s$) is an increasing function of $\lambda_B$ as also shown in Fig. \ref{p_f_vs_lambda_B_1}. That is, increasing the number of BSs allows the network to schedule more users at the same time slot which in turn increases the number of concurrent cellular transmissions that the D2D transmitters can harvest from. 

Overall, it can be seen in Figs. \ref{p_f_vs_lambda_B_2} and \ref{p_f_vs_lambda_B_1} that $\lambda_B$ balances the trade-off between $p_f$ and $p_s$. For instance, it is not helpful to have a very high probability of finding the D2D channel free by decreasing the intensity of BSs while having a very low chance to harvest sufficient energy, and vice versa. For the downlink scenario, it can be seen that as $\lambda_B$ increases, the transmission probability of a D2D transmitter increases up to a maximum value, then it starts to decrease. The behavior can be explained as follows: for networks with low intensity of BSs, the effect of harvesting sufficient energy  dominates the transmission probability and improves it despite the degradation in $p_f$. However, as $\lambda_B$ increases, the probability of harvesting sufficient energy $p_s$ saturates and the effect of the degradation in $p_f$ starts to dominate the D2D transmission probability. Hence, the transmission probability starts to decrease. For the uplink scenario, since $p_f$ is constant, the overall performance simply follows the same trend of $p_s$ as shown in Fig. \ref{p_f_vs_lambda_B_2}. Furthermore, this figure shows that the PSA policy always outperforms the RSA policy for any value of $\lambda_B$. This result is intuitive since the PSA policy offers a better performance in terms of $p_f$ by avoiding the use of $c_d$ as much as possible when compared to RSA as explained in the Section \ref{sec:sysModel}. 

It is also observed that the using an uplink channel for D2D communication is more beneficial than using a downlink channel. This is due to the difference in transmit powers of the BSs in downlink and the cellular users in uplink. This means that, under the same sensing threshold, the D2D transmitter needs to avoid less amount of interference when using an uplink channel when compared to a downlink channel, which increases $p_f$. Note that, from \textbf{Lemma \ref{lem_lambda_ref}}, $\lambda_B^{\rm ref} = 1.42$ BS/km$^2$ in this case which coincides with the results presented in Fig. \ref{p_f_vs_lambda_B_2}.

\subsection{Outage Probability for D2D users}

We now discuss the $\mathsf{SINR}$ outage probability as well as the overall outage probability for D2D users and show the effect of different network parameters (e.g., $\lambda_B$, $\gamma$, $|\mathcal{C}|$) on these important performance metrics.

\begin{figure}[!t]
\centering
\includegraphics[width=8.7cm]{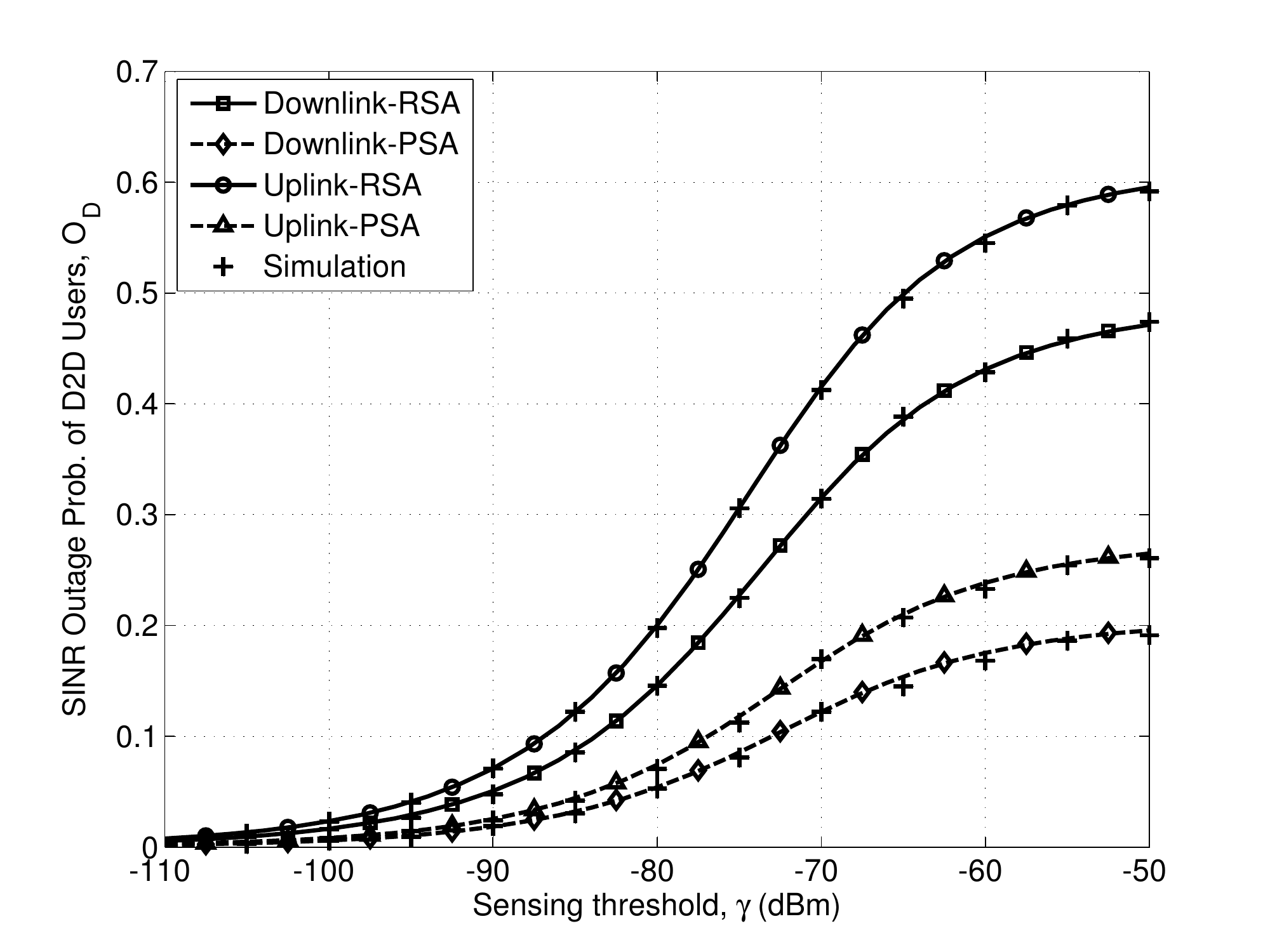}
\caption{The $\mathsf{SINR}$ outage probability $\mathsf{O}_D$ for D2D transmissions vs. spectrum sensing threshold (in dBm) when $c_d$ is a downlink or an uplink channel. The network parameters are $|\mathcal{C}| = 15$ channels, $\rho_d = -70$ dBm, and $d_o = 10$ m. The results are shown for both  the RSA and PSA policies.}
\label{D2D_SINR_vs_gamma}
\end{figure}

Fig. \ref{D2D_SINR_vs_gamma} depicts the $\mathsf{SINR}$ outage probability for D2D users as a function of the spectrum sensing threshold for the different scenarios. As stated in \textbf{Theorem \ref{thm_O_D}}, for all scenarios, decreasing $\gamma$ improves the performance of the $\mathsf{SINR}$ outage probability by offering more protection for the D2D transmissions. This result highlights the importance of carefully choosing the spectrum sensing threshold after considering its effect on both the transmission probability and the $\mathsf{SINR}$ outage probability. In Fig. \ref{D2D_SINR_vs_gamma}, it can also be seen that the PSA policy offers a better coverage\footnote{Coverage is defined as the complimentary event of the outage, i.e., the coverage probability is equal to $1-\mathsf{O}$.} compared to the RSA policy for all values of $\gamma$ irrespective of whether $c_d$ is a downlink or an uplink channel. 
For example, when $\gamma = -60$ dBm, the outage reduces from $43\%$ to $17\%$ dBm when $c_d$ is a downlink channel and from $55\%$ to $24\%$ dBm when $c_d$ is an uplink channel. The PSA policy reduces the probability of cellular users to access the D2D channel; hence, it reduces the number of active interferers on this channel, and consequently, improves the $\mathsf{SINR}$. In order to compare the downlink and uplink channel cases for $c_d$, we recall Fig. \ref{p_f_vs_lambda_B_2} in which we can see that for the chosen simulation parameters (i.e., $\lambda_B = 1$ BS/km$^2$), the transmission probabilities of the cellular transmitters for the uplink channel case are higher than those for the downlink channel case (since $\lambda_B^{\rm ref} = 1.42$ BS/km$^2$ from \textbf{Lemma \ref{lem_lambda_ref}}). Hence, when $c_d$ is an uplink channel, the intensity of interferers  is higher and the $\mathsf{SINR}$ outage is higher when compared to the case when $c_d$ is a downlink channel, cf. Fig. \ref{D2D_SINR_vs_gamma}. Note that, if we increase $\lambda_B$ (i.e., beyond $\lambda_B^{\rm ref}$), the uplink case will outperform the downlink case. This is for the same reason as that for Fig. \ref{p_f_vs_lambda_B_2}. That is, as $\lambda_B$ increases, the transmit power of cellular users decreases, and consequently, the aggregate interference decreases when compared to the downlink case in which the transmit power is fixed.

\begin{figure}[!t]
\centering
\includegraphics[width=8.7cm]{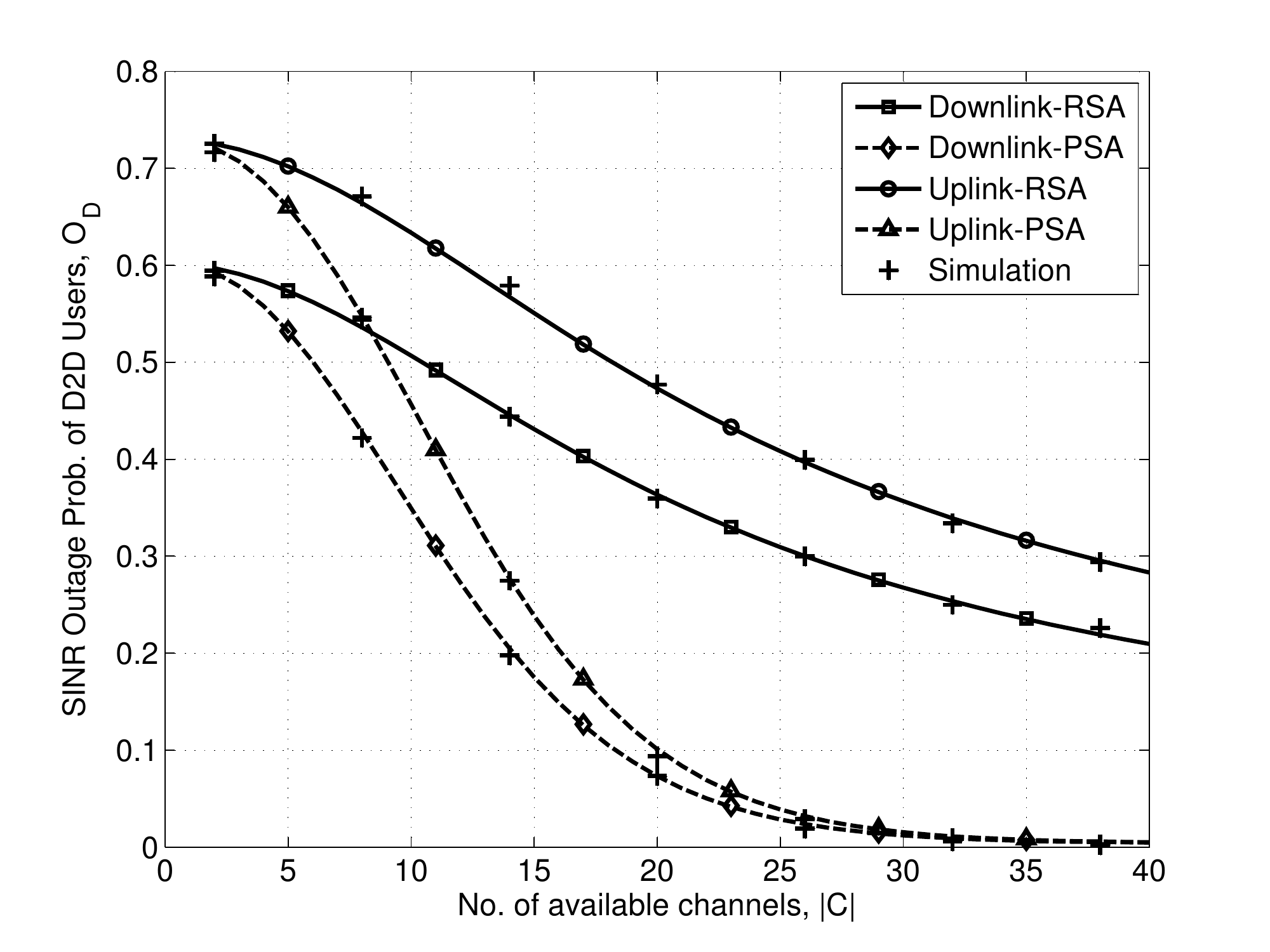}
\caption{The $\mathsf{SINR}$ outage probability $\mathsf{O}_D$ for D2D transmissions vs. the number of available channels when $c_d$ is a downlink or an uplink channel. The network parameters are $\rho_d = -70$ dBm, $d_o = 10$ m, and $\gamma = -60$ dBm. The results are shown for both  the RSA and PSA policies.}
\label{D2D_SINR_vs_C}
\end{figure}

Fig. \ref{D2D_SINR_vs_C} shows the effect of varying the number of available channels (i.e., downlink or uplink) on the $\mathsf{SINR}$ outage probability of D2D users for different scenarios. It can be seen that increasing $|\mathcal{C}|$ can improve the performance of both the spectrum access policies and both the uplink and downlink channel cases for $c_d$. Note that the PSA policy outperforms the RSA policy where it can achieve very low $\mathsf{SINR}$ outage probability for a relatively small number of channels. For example, in an uplink channel scenario, using only $11$ channels provides an outage of less than $0.3$ for the PSA policy. On the other hand, for the RSA policy, at least $25$ channels are required to achieve similar outage performance.

\begin{figure}[!t]
\centering
\includegraphics[width=8.7cm]{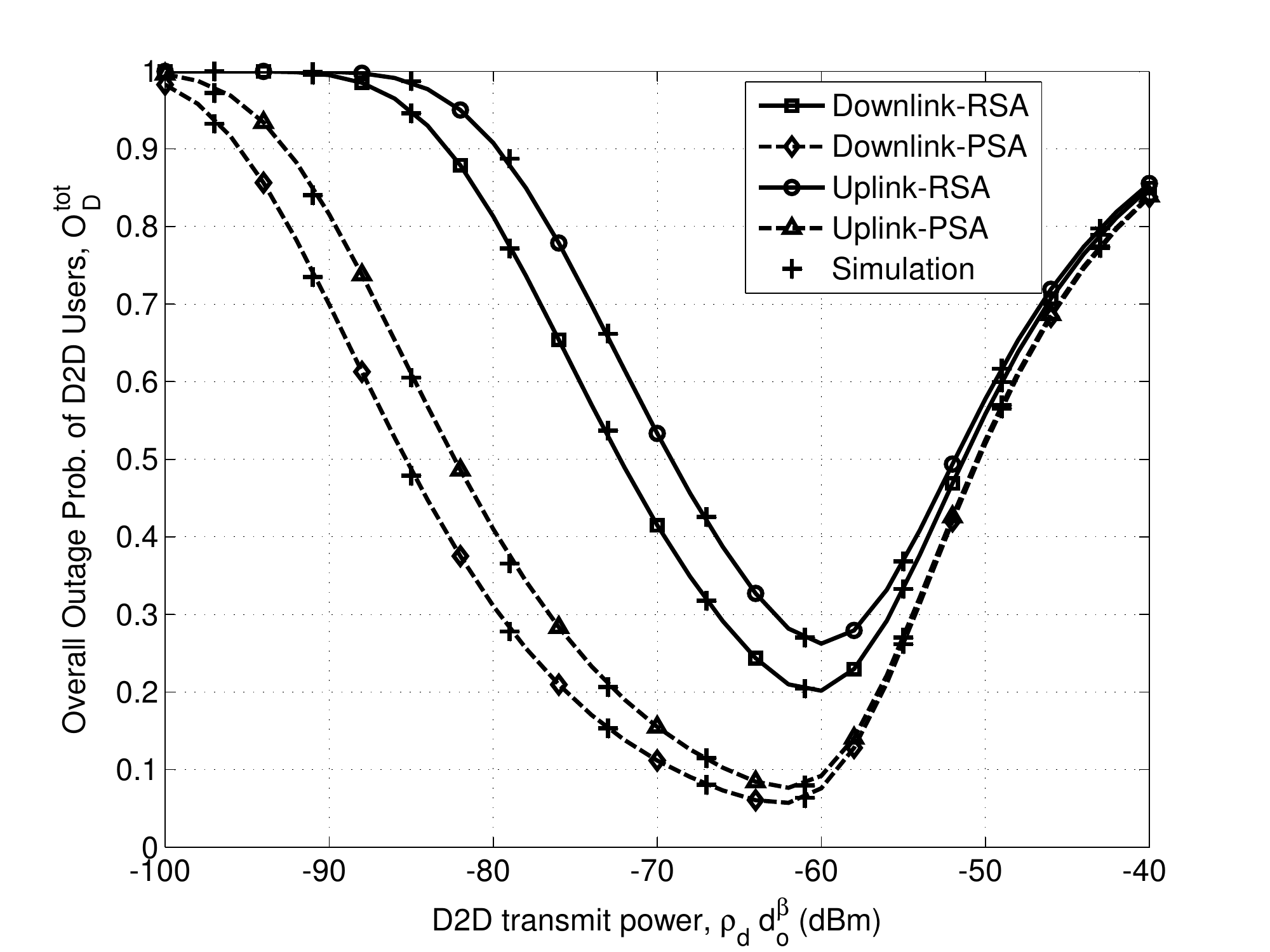}
\caption{The overall outage probability $\mathsf{O}_D^{\rm tot}$  for D2D transmissions vs. the receiver sensitivity (in dBm) when $c_d$ is a downlink and an uplink channel. The network parameters are $\lambda_U = 5\lambda_B$, $\gamma = -60$ dBm and $|\mathcal{C}| = 10$ channels. The results are shown for both  the RSA and PSA policies.}
\label{D2D_vs_rho_d}
\end{figure}

Fig. \ref{D2D_vs_rho_d} shows variations in the overall outage probability for D2D users (as given in (\ref{eq:O_D_tot})) with the sensitivity of the D2D receiver ($\rho_d$) for all scenarios (or equivalently, the transmit power of a D2D user when $d_o$ is fixed). We observe that there is an optimal value of $\rho_d$ which minimizes the overall outage probability. Starting from the optimal point, when $\rho_d$ decreases, $p_s$ approaches 1;  however, the $\mathsf{SINR}$ decreases due to decrease in the received power of the useful signal compared to the aggregate interference. Therefore, the $\mathsf{SINR}$ outage probability ($\mathsf{O}_D$) dominates and the overall outage probability increases. When $\rho_d$ increases from the optimal point, the $\mathsf{SINR}$ outage improves; however,  $p_s$ decreases. In this case $p_s$ dominates the performance and the overall outage probability increases. The same observations can be made when comparing the different scenarios as in Figs. \ref{p_f_vs_lambda_B_2}-\ref{D2D_SINR_vs_C}.

\subsection{Outage Probability for Cellular Users}
\begin{figure}[!t]
\centering
\includegraphics[width=8.7cm]{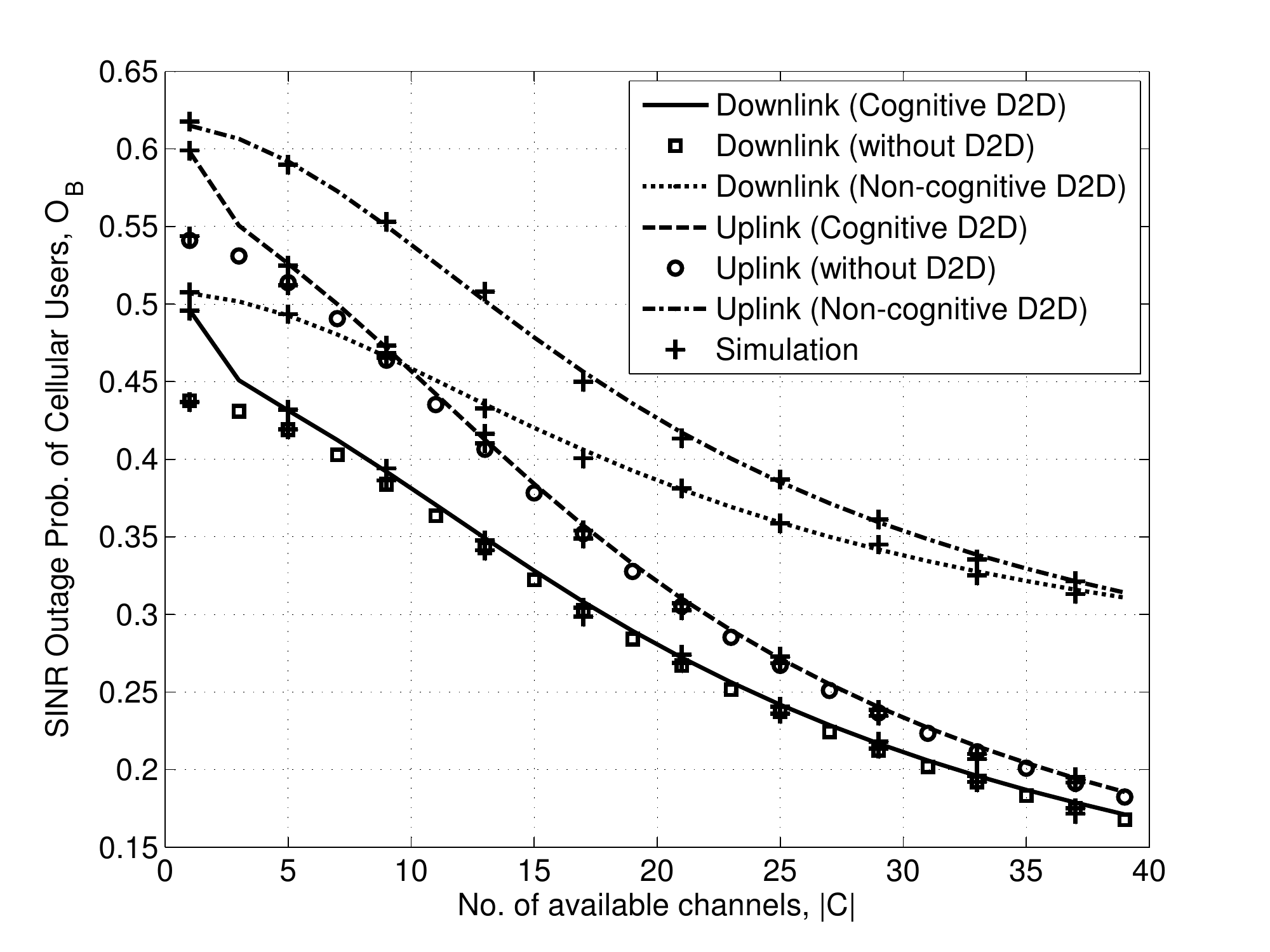}
\caption{The average $\mathsf{SINR}$ outage probability $\mathsf{O}_B$ for cellular transmissions vs. the number of available channels when $c_d$ is a downlink or an uplink channel. The network parameters are $\rho_d = -60$ dBm, $\beta = 4$, $d_o = 15$ m, and $\gamma = -60$ dBm. The results are shown for the RSA policy.}
\label{cellular_SINR_vs_C}
\end{figure}

In order to show the effect of cognition and D2D channel selection on the performance of cellular users, Fig. \ref{cellular_SINR_vs_C} shows the average $\mathsf{SINR}$ outage probability in presence of D2D communication with and without cognition. Note that $\mathsf{O}_B$ is averaged over all channels. In this comparison, we consider the performance of the case without D2D transmission as our reference scenario (i.e., shown by curves with circle and square markers and no lines). In addition, we consider a worst-case scenario when all D2D transmitters have sufficient energy to perform channel inversion power control, i.e., $p_s = 1$. We can see that  D2D communication without cognition degrades the $\mathsf{SINR}$ outage probability (i.e., shown by curves with circle and square markers and solid lines). From a cellular user's perspective, this  is expected as new interferers (i.e., D2D transmitters) are added to the system. With cognitive D2D transmission, the interference caused to the cellular users can be mitigated (i.e., shown by curves with dashed lines and no markers). Note that, cognitive D2D transmissions protect the cellular transmissions by controlling the intensity of active D2D transmitters.  Note also that the same results and observations in Fig. \ref{cellular_SINR_vs_C} hold for the PSA policy with a very slight difference in the performance as can be noticed from (\ref{eq:q_d_reduction}) and (\ref{eq:q_c_increase}).

\section{Conclusion}
\label{sec:conc}
We have presented a novel model for cognitive D2D communication using RF energy harvesting from the ambient interference in a multi-channel downlink-uplink cellular network. For coexistence of the underlaying D2D transmissions in downlink and uplink channels, we have proposed two different spectrum access policies for the cellular network, namely, random and prioritized access policies. We have used stochastic geometry to provide a complete framework to model, analyze, and evaluate the performance of the proposed system in terms of transmission probability and $\mathsf{SINR}$ outage probabilities for D2D and cellular users.  
Under the same network setup, the prioritized spectrum access method outperforms the  random spectrum access method for all considered performance metrics for the D2D users. Furthermore, for the cellular users, the effect of the prioritized spectrum access policy adopted by the BSs has been observed to be negligible compared to the random access policy. 
In addition, we have shown that, while uplink channels are preferable to downlink channels for D2D transmissions in dense cellular networks, downlink channels provide better performance in cellular networks with low density of BSs.
We conclude that by carefully tuning the network design parameters, energy harvesting can be used along with cognitive D2D transmission to provide an acceptable quality of service performance of D2D communication without significantly affecting the performance of cellular communication.

\appendices
\renewcommand{\thesubsection}{\thesection -\Roman{subsection}}
\def\thesubsectiondis{\Roman{subsection}.} 

\section{Proof of Access Probabilities}
\subsection{Proof of Lemma \ref{lem_q_f}}
\label{lem_q_f_proof}
Since the number of used channels by a BS is $\min\{ N_u, |\mathcal{C}| \}$, by conditioning on the number of users served by a generic BS such that $N_u = n$, the conditional probability that a BS assigns at least one channel for a generic cellular user is given by
\begin{align}
q_{f|n} = 
\begin{cases}
	1, & 0 \leq n \leq |\mathcal{C}|\\	
	\frac{|\mathcal{C}|}{n}                  , & n > |\mathcal{C}|.
\end{cases}
\end{align}   
Therefore, the unconditional probability can be obtained as 
\begin{align}
q_{f} &= \sum_{n=0}^\infty q_{f|n} \mathbb{P} \{ N_u = n \} \nonumber\\
&= \sum_{n=0}^{|\mathcal{C}|} \mathbb{P} \{ N_u = n \} + \sum_{n=|\mathcal{C}|+1}^{\infty} \frac{|\mathcal{C}|}{n} \mathbb{P} \{ N_u = n \}.
\end{align}
Using the fact that $\sum_{n=0}^\infty \mathbb{P} \{ N_u = n \} = 1$, the expression of $q_f$ in (\ref{eq:q_f}) can be obtained.

\subsection{Proof of Lemma \ref{lem_RSA}}
\label{lem_RSA_proof}

By conditioning on the number of users served by a generic BS such that $N_u = n$, when using RSA, the conditional probability that a BS uses any channel $c_i \in \mathcal{C}$ is
\begin{align}
q_{c|n} = \frac{\binom{|\mathcal{C}| - 1}{n - 1}}{\binom{|\mathcal{C}|}{n}} =  
\begin{cases}
	\frac{n}{|\mathcal{C}| }, & n < |\mathcal{C}|\\	
	1                                     , & n \geq |\mathcal{C}|.
\end{cases}
\end{align}   
Therefore, the unconditional probability can be obtained as 
\begin{align}
q_{c} &= \sum_{n=0}^\infty q_{c|n} \mathbb{P} \{ N_u = n \} \label{eq:q_c} \nonumber\\
&= \sum_{n=0}^{|\mathcal{C}|-1} \frac{n}{|\mathcal{C}|}  \mathbb{P} \{ N_u = n \} + \sum_{n=|\mathcal{C}| }^{\infty} \mathbb{P} \{ N_u = n \}.
\end{align}
Using $\sum_{n=0}^\infty \mathbb{P} \{ N_u = n \} = 1$ and since all channels are used with the same probability when adopting the RSA policy, the expressions of $q_c^{\text{RSA}}$ and $q_d^{\text{RSA}}$ in (\ref{eq:q_c_RSA}) can be easily verified.

\subsection{Proof of Lemma \ref{lem_PSA}}
\label{lem_PSA_proof}
When using PSA, each BS randomly and independently uses any channel $c_i\in \mathcal{C} \setminus \{c_d\}$ with the same probability. By following the same proof of \textbf{Lemma \ref{lem_RSA}} while using the number of available channels to be $|\mathcal{C}| - 1$ instead of $|\mathcal{C}| $, the expression for the unconditional probability $q_c$ in (\ref{eq:q_c_PSA}) can be easily obtained. Note that, the last term of the summation is zero when $n=|\mathcal{C}| - 1$.

The channel $c_d$ is used only when the BS has no other channels to use, i.e., when the number of cellular users associated to that BS is greater then $|\mathcal{C}|-1$. Hence,
\begin{align}
q_d &= \mathbb{P} \{ N_u \geq |\mathcal{C}| \}
\end{align}
which is equivalent to (\ref{eq:q_d_PSA}).

\section{Proof of Lemma \ref{lem_p_f}}
\label{lem_p_f_proof}
Here we present the proof when $c_d$ is a downlink channel where the same proof follows for the uplink case. For a generic D2D transmitter in the origin $(0,0) \in \mathbb{R}^2$, firstly, we obtain the distribution of the number of BSs that use $c_d$ in the protection region $\mathcal{R}(\gamma)$ defined in (\ref{eq:cog_area_DL}). Using the thinning operation on the PPP of the BSs $\mathbf{\Phi}_B$, the BSs that use the D2D channel $c_d$ can be modeled by a homogenous PPP $\mathbf{\tilde{\Phi}}_B$ with intensity $\lambda_B' = q_d \lambda_B$. We define $N_{\mathcal{R}(\gamma)}$ as the number of BSs that use $c_d$ in the protection region $\mathcal{R}(\gamma)$.

Since the BSs that use the D2D channel constitute a PPP, the number of BSs in $\mathcal{R}(\gamma)$ from $\mathbf{\tilde{\Phi}}_B$ is a Poisson-distributed random variable with parameter $\lambda_B' \times \nu(\mathcal{R}(\gamma))$, i.e., $N_{\mathcal{R}(\gamma)}\sim \text{Poisson}(\lambda_B' \times \nu(\mathcal{R}(\gamma))$, where
\begin{align}
\nu(\mathcal{R}(\gamma)) &= \mathbb{E}\left[\pi {r}_{P}^2 \right] = \pi \left(\frac{P_B}{ \gamma}\right)^{\frac{2}{\alpha}} \Gamma\left(\frac{\alpha+2}{\alpha}\right).
\end{align}
By definition, we obtain
\begin{align}
p_{f} &= \mathbb{P}[N_{\mathcal{R}(\gamma)} = 0] = \exp[-\lambda_B'~ \nu(\mathcal{R}(\gamma))].
\end{align}

For the uplink case, $p_f$ follows by using the $\frac{2}{\alpha}$-th moment of the transmit power of a cellular user, i.e.,  $\mathbb{E} \left[ P_u^\frac{2}{\alpha} \right] = \frac{\rho_b^\frac{2}{\alpha}}{\pi \lambda_B}$ when using channel inversion uplink power control \cite{sakrup}.

\section{Proof of Lemma \ref{lem_p_s}}
\label{lem_p_s_proof}
We find the distribution of the aggregate interference power received at the D2D transmitter located at the origin by calculating its Laplace transform. That is,
\begin{align}
\mathcal{L}_{P_{\rm H}}\left( s \right) &= \mathcal{L}_{P_{\rm H}}^{\rm D}\left( s \right)  \mathcal{L}_{P_{\rm H}}^{\rm U}\left( s \right) \label{eq:lap}
\end{align}
where the superscripts ${\rm D}$ and ${\rm U}$ refer to the downlink and uplink channels as presented in (\ref{eq:P_h}) by the first and second terms, respectively.

Firstly, we start by obtaining $\mathcal{L}_{P_{\rm H}}^{\rm D}\left( s \right)$ for the downlink subset of channels $\mathcal{C}_D$  as 
\begin{align}
\mathcal{L}_{P_{\rm H}}^{\rm D}\left( s \right) &= \mathbb{E}_{\mathbf{\tilde{\Phi}}_B,\{h\}} \left[ \exp \left[ -s a P_B \sum_{x \in \mathbf{\tilde{\Phi}}_B} \sum_{c \in \mathcal{C_D}} h R_i^{-\alpha}  \right] \right] \nonumber\\
&\stackrel{(a)}{=} \mathbb{E}_{\mathbf{\tilde{\Phi}}_B} \left[\prod_{x_i \in \mathbf{\tilde{\Phi}}_B} \left( \frac{1}{1+s a P_B R_i^{-\alpha}} \right)^{|\mathcal{C}_D|} \right]
\label{eq:Lap_P_h_D}
\end{align}
where $(a)$ follows because of the independence assumption of Rayleigh fading and by using the moment generating function of an exponential random variable. By using the probability generating functional (PGFL) of PPP, we obtain
\begin{align}
\mathcal{L}_{P_{\rm H}}^{\rm D}\left( s \right) 	&= \Scale[1.0]{\exp \left[ -2 \pi q_c^{\rm D} \lambda_B  {\displaystyle\int_{0}^\infty} \! \! \left( 1 - \left(\frac{1}{1+s a P_B r^{-\alpha}} \right)^{\! |\mathcal{C}_D|} \right) r \text{d}r \right]} \nonumber\\
&\stackrel{(b)}{=} \exp \left[ - \kappa_1 s^{\frac{2}{\alpha}} \right] \label{eq:lapD}
\end{align}
where $\kappa_1$ is defined in (\ref{eq:kappa_1}) and $q_c^{\rm D}$ is given in (\ref{eq:q_c_RSA}) and (\ref{eq:q_c_PSA}) for the RSA and PSA policies, respectively, by using $|\mathcal{C}_D|$. In addition, $(b)$ follows by replacing $u=\frac{1}{1+s P_B r^{-\alpha}}$.

Unlike the downlink network in which a BS can establish up to $|\mathcal{C}_D|$ transmissions over many channel, each user in the uplink network can establish only one connection using only one of the $\mathcal{C}_U$ channels. Therefore, we define a point process $\mathbf{\Phi'}_U = \bigcup\limits_{c \in \mathcal{C}_U} \mathbf{\tilde{\Phi}}_U(c)$ with intensity $q_c^{\rm U} |\mathcal{C}_U| \lambda_B$ that represents all the transmitting cellular users. Note that the intensity of this point process is calculated based on the fact that each BS receives only up to $|\mathcal{C}_U|$ transmissions at any time. Hence, the average number of uplink transmissions received by a BS is equal to $\sum \limits_{n=0}^\infty \min\{n,|\mathcal{C}_U|\} \mathbb{P}\{N_u=n\} = q_c^{\rm U} |\mathcal{C}_U|$ where $q_c^{\rm U}$ is obtained by \textbf{Lemma} \ref{lem_RSA} (or \ref{lem_PSA}) after replacing $|\mathcal{C}|$ by $|\mathcal{C}_U|$. Now, we obtain $\mathcal{L}_{P_{\rm H}}^{\rm U}\left( s \right)$ for the uplink subset of channels $\mathcal{C}_U$ as follows:
\begin{align}
\mathcal{L}_{P_{\rm H}}^{\rm U}\left( s \right) &= \mathbb{E}_{\mathbf{\Phi'}_U} \left[\prod_{u_i \in \mathbf{\Phi'}_U} \mathbb{E}_{P_u} \left[\frac{1}{1+s a P_u R_i^{-\alpha}}  \right] \right] \nonumber\\
&\stackrel{(c)}{=} \exp \left[ - \kappa_2 s^{\frac{2}{\alpha}} \right] \label{eq:lapU}
\end{align}
where $\kappa_2$ is defined in (\ref{eq:kappa_2}) and $(c)$ follows by using PGFL of PPP and the $m$-th moment of the user's transmit power. Note that $\mathbf{\Phi'}_U$ is not a PPP where this assumption will be validated by simulations in Section \ref{sec:results}.

Using (\ref{eq:lap}), (\ref{eq:lapD}), and (\ref{eq:lapU}), we obtain the Laplace transform of the aggregate received interference as: $\mathcal{L}_{P_{\rm H}}\left( s \right) = \exp [ - \kappa_3  s^{\frac{2}{\alpha}} ]$ where $\kappa_3 = \kappa_1+\kappa_2$.

Now, we derive the cumulative  distribution function ({\em cdf}) of the harvested power using the inverse Laplace transform method. Specifically, we use the Bromwich inversion theorem with the modified contour defined in \cite[Chapter 2]{brown}. Hence,
\begin{align}
F_{P_{\rm H}}(t) &\stackrel{(d)}{=} \frac{1}{2\pi i}\lim_{T\to\infty}\int_{\gamma-iT}^{\gamma+iT} \exp\left[st - \kappa_3 s^{\frac{2}{\alpha}} \right]\frac{\text{d}s}{s} \nonumber\\
&\stackrel{(e)}{=} 1 - \frac{1}{2\pi i} \int_0^\infty e^{- u t} \left(e^{\kappa_3 (-u)^\frac{2}{\alpha}} - e^{-\kappa_3 (-u)^\frac{2}{\alpha}}\right) \frac{\text{d}u}{u}  \label{eq:F_P_h_1}
\end{align}
where $(e)$ follows because the integrand in $(d)$ has a branch point at the origin. Then, according to the definition of $p_s$ and by using the expression for $P_D$, (\ref{eq:p_s_4}) can be easily verified. For the detailed derivations of $(b)$, $(c)$, and $(e)$, refer to the auxiliary appendix in \cite{page}.

\section{Proof of Theorem \ref{thm_O_D}}
\label{thm_O_D_proof}
By definition, we obtain the complementary {\em cdf} of the $\mathsf{SINR}_D$ as
\begin{align}
\mathbb{P} \left[ \mathsf{SINR}_D > \tau \right] &= \mathbb{P} \left[ h_{y_0} > \tau \frac{I_B + I_D + \sigma_z^2}{ \rho_d} \right]	\nonumber\\
&\stackrel{(f)}{=} \exp \left[\frac{- \tau \sigma_z^2}{\rho_d} \right] \mathcal{L}_{I_B}\left( \frac{\tau }{\rho_d} \right) \mathcal{L}_{I_D}\left( \frac{\tau }{\rho_d } \right) \label{eq:CCDF_D}
\end{align}
where $(f)$ follows because the channel fading coefficient $h_{y_0} \sim \Exp(1)$ while $\mathcal{L}_{I_B}$ and $\mathcal{L}_{I_D}$ are the Laplace transforms of the aggregate interference resulting from, respectively, the cellular and D2D transmissions, evaluated at $\frac{\tau }{\rho_d}$.

For the Laplace transform of $I_D$, by following the proof of \textbf{Lemma \ref{lem_p_s}} in \textbf{Appendix \ref{lem_p_s_proof}} when $|\mathcal{C}|=1$, we obtain
\begin{align}
\mathcal{L}_{I_D}\left( \frac{\tau }{\rho_d} \right) &=\exp \left[ - \frac{2 \pi^2 d_0^2 p_s p_{f} \lambda_D}{\beta \sin\left( \frac{2\pi}{\beta} \right)} \tau^{\frac{2}{\beta}} \right]. \label{eq:L_DD}
\end{align}

For $I_B$, since the D2D transmitters use spectrum sensing before transmission, the nearest interfering macro BS (or cellular user) when $c_d$ is a downlink (or uplink) channel is at least at a distance of $\bar{r}_P$ from the intended D2D receiver where $\bar{r}_P$ is given in (\ref{eq:r_P_DL}) (or (\ref{eq:r_P_UL})). Note that in order to protect the D2D transmissions, the protection region should be centered around the receiver rather than the transmitter. However, for  simplicity, we assume that the protection region is centered around the D2D transmitter while the maximum separation between the transmitter and the receiver $d_o << \bar{r}_{P}$ so that the D2D receiver is well protected by the cognition performed by its corresponding transmitter. Hence, by following (\ref{eq:Lap_P_h_D}) we obtain
\begin{align}
\mathcal{L}_{I_B}\left( s \right)	&= \exp \left[ -2 \pi q_d \lambda_B \int_{\bar{r}_P}^\infty \mathbb{E}_P \left[\frac{r}{1+\frac{1}{s P}  r^{\alpha}}\right] \text{d}r \right]
\end{align}
where $P$ is the transmit power $P_B$ of a BS, when $c_d$ is a downlink channel, otherwise, $P$ is the transmit power $P_u$ of a cellular user.

Now, by replacing $\frac{1}{s P} r^{\alpha}$ with $u^{\alpha}$ and $\bar{r}_P=\left(\frac{P}{\gamma}\right)^{\frac{1}{\alpha}} \Gamma\left(1+\frac{1}{\alpha}\right)$, we obtain
\begin{align}
\mathcal{L}_{I_B}\left( s \right) &= \exp \! \left[  -2 \pi q_d \lambda_B \mathbb{E}[P^{\frac{2}{\alpha}}] s^{\frac{2}{\alpha}} \! \! \int_{\left(\frac{1}{\gamma s}\right)^{\frac{1}{\alpha}} \Gamma\left(1+\frac{1}{\alpha}\right)}^\infty \! \!  \frac{u}{1+u^{\alpha}} \text{d}u \right]\! \! .\label{eq:L_BD}
\end{align}

By combining (\ref{eq:CCDF_D}), (\ref{eq:L_DD}), and (\ref{eq:L_BD}),  we obtain the $\mathsf{SINR}$ outage probability $\mathcal{O}_D$ of a typical D2D receiver.

\section{Proof of Theorem \ref{thm_O_B}}
\label{thm_O_B_proof}
To calculate the outage of cellular downlink and uplink transmissions, we use an approach similar to that in the proof of \textbf{Theorem \ref{thm_O_D}}. Note that for $I_D$, the interfering D2D transmitters can be arbitrarily close to the tagged receiver (macro BS in uplink or cellular user in downlink) and there is no protection region. For the cellular interference in downlink, since each user associates with the closest BS, no interfering BS can be closer to the tagged user than the serving BS. For the interference in the uplink, with the channel inversion power control, the closest interferer to the tagged BS is at lease at a distance $(\frac{P_u}{\rho_b})^{\frac{1}{\alpha}}$. Based on the aforementioned facts, the Laplace transform of $I_D$ is the same as in (\ref{eq:L_DD}) and evaluated at $\frac{\tau}{P_B r^{-\alpha}}$ for downlink and at $\frac{\tau}{\rho_b}$ for uplink. On the other hand, that of $I_B$ can be obtained by following \textbf{Appendices \ref{lem_p_s_proof}} and \ref{thm_O_D_proof} while using the protection radius defined earlier.

Then, knowing that the distribution of the distance between a generic cellular user and its serving BS is Rayleigh, i.e., $f_R(r) = 2 \pi \lambda_B ~r \exp \left[ -\pi \lambda_B r^2 \right]$, using (\ref{eq:outage_gen_B}), the outage probability can be easily verified.

\begin{IEEEbiography} [{\includegraphics[width=1in,height=1.25in,clip,keepaspectratio]{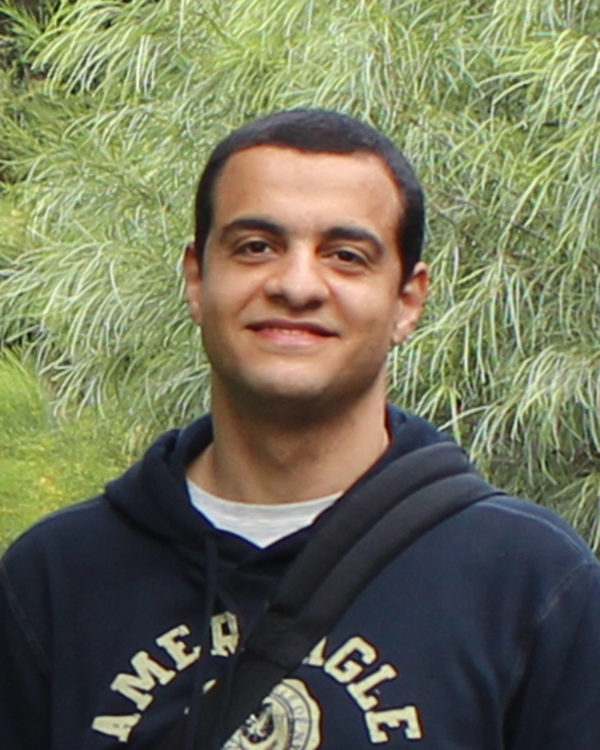}}]
{Ahmed H. Sakr} (S'12) is a Ph.D. candidate in the Department of Electrical and Computer Engineering, University of Manitoba, Canada.  He received the B.Sc. (2002-2007) and M.Sc. (2010-2012) degrees both in Electronics and Communications Engineering from Tanta University, Tanta, Egypt, and Egypt-Japan University of Science and Technology (E-JUST), Alexandria, Egypt, respectively. For his academic excellence, he has received several academic awards including the Manitoba Graduate Scholarship (MGS) in 2014, Edward R. Toporeck Graduate Fellowship in Engineering in 2014, the Graduate Enhancement of Tri-Council Stipends (GETS) in 2013, and Egyptian Ministry of Higher Education Excellence Scholarship in 2010-2012. Ahmed has been a member in the technical program committee and a reviewer in several IEEE journals and conferences. His current research interests include statistical modeling of wireless networks, resource allocation in multi-tier cellular networks, and green communications. (http://home.cc.umanitoba.ca/\texttildelow{}sakra.html)
\end{IEEEbiography}

\begin{IEEEbiography} [{\includegraphics[width=1in,height=1.25in,clip,keepaspectratio]{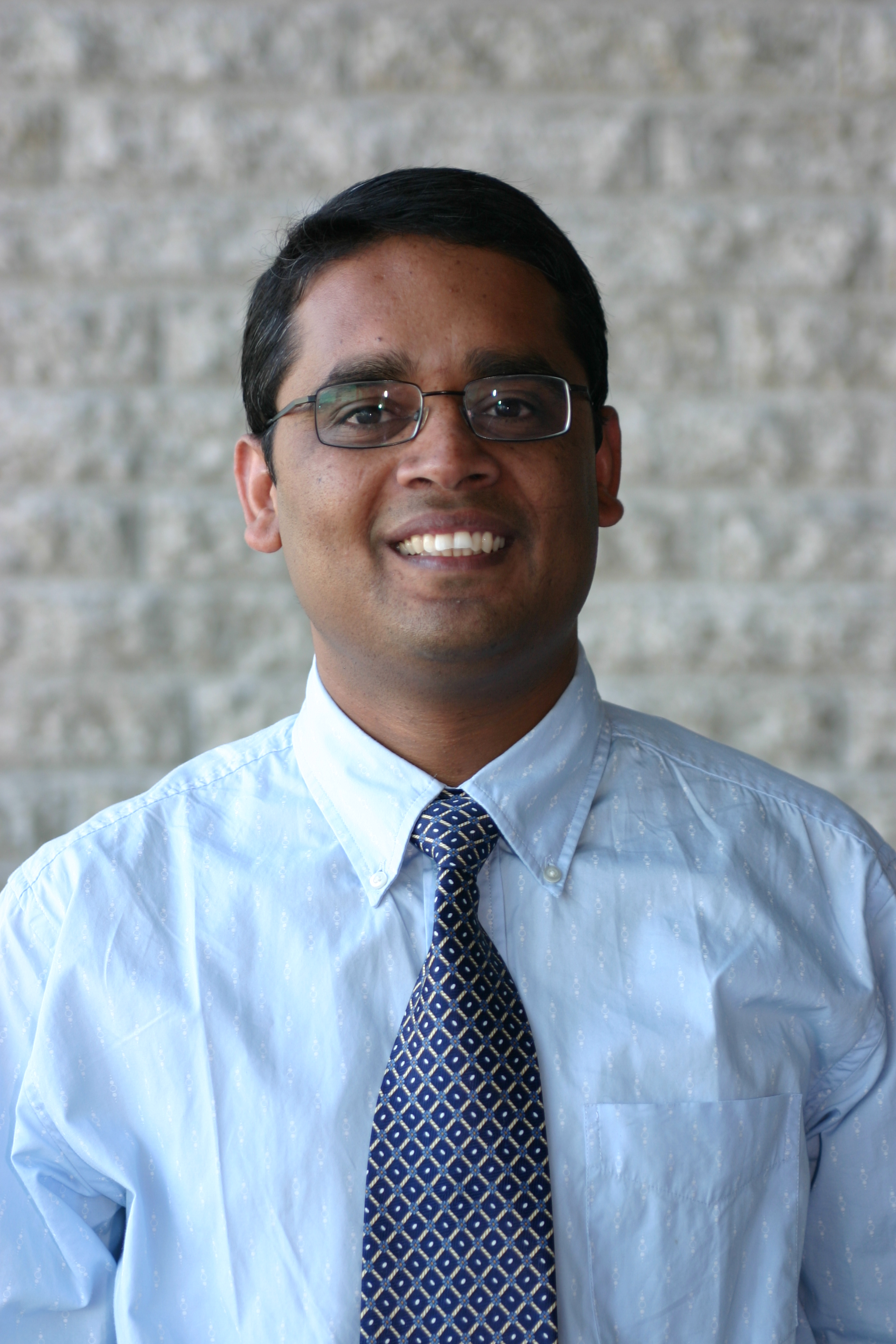}}]
{Ekram Hossain} (F'15)
is a Professor (since March 2010) in the Department of Electrical and Computer Engineering at University of Manitoba, Winnipeg, Canada. He received his Ph.D. in Electrical Engineering from University of Victoria, Canada, in 2001. Dr. Hossain's current research interests include design, analysis, and optimization of wireless/mobile communications networks, cognitive radio systems, and network economics. He has authored/edited several books in these areas (http://home.cc.umanitoba.ca/\texttildelow{}hossaina). Dr. Hossain serves as the Editor-in-Chief for the {\em IEEE Communications Surveys and Tutorials} and an Editor for {\em IEEE Wireless Communications}. Also, he is a member of the IEEE Press Editorial Board. Previously, he served as the Area Editor for the {\em IEEE Transactions on Wireless Communications} in the area of ``Resource Management and Multiple Access'' from 2009-2011, an Editor for the {\em IEEE Transactions on Mobile Computing} from 2007-2012, and an Editor for the {\em IEEE Journal on Selected Areas in Communications - Cognitive Radio Series} from 2011-2014. Dr. Hossain has won several research awards including the University of Manitoba Merit Award in 2010 and 2014 (for Research and Scholarly Activities), the 2011 IEEE Communications Society Fred Ellersick Prize Paper Award, and the IEEE Wireless Communications and Networking Conference 2012 (WCNC'12) Best Paper Award. He is a Distinguished Lecturer of the IEEE Communications Society (2012-2015). Dr. Hossain is a registered Professional Engineer in the province of Manitoba, Canada.
\end{IEEEbiography}

\end{document}